%% file: ms.tex
\documentclass[twocolumn,twocolappendix]{aastex7}
\usepackage{showyourwork}
\usepackage{multirow}
\usepackage{gensymb}
\usepackage{datetime}
\usepackage{CJK}
\usepackage[T1]{fontenc}
\usepackage{amsmath}
\usepackage{hyperref}
\usepackage[capitalize,noabbrev]{cleveref}
\yyyymmdddate
\usepackage{mathtools}
\usepackage{savesym}
\let\tablenum\relax
\usepackage{siunitx}[=v2]
\usepackage{chemformula}
\let\oldenddeluxetable\endxdeluxetable
\let\olddeluxetable\xdeluxetable
\makeatletter
\renewenvironment{xdeluxetable}[1]{
\olddeluxetable{[#1]}
\def\pt@format{\string#1}%
}{\oldenddeluxetable}
\makeatother

\DeclareSIUnit{\mag}{mag}
\DeclareSIUnit{\mas}{mas}
\DeclareSIUnit{\electron}{e^-}
\DeclareSIUnit{\pixel}{px}
\DeclareSIUnit{\adu}{adu}
\DeclareSIUnit{\year}{yr}
\DeclareSIUnit{\parsec}{pc}
\DeclareSIUnit{\au}{au}
\DeclareSIUnit{\Msun}{M_\odot}
\DeclareSIUnit{\Mjup}{M_J}

\received{2025/05/20}
\revised{2025/09/17}
\accepted{2025/09/17}
\submitjournal{AJ}

\usepackage{etoolbox}
\makeatletter
\pretocmd\@sect{\def\@currentcounter{#1}}{}{\fail}
\makeatother

\begin{document}
\begin{CJK*}{UTF8}{gbsn}

\title{Dynamical Analysis of the HD 169142 Planet-Forming Disk: Twelve Years of High-Contrast Polarimetry}
\shorttitle{Dynamical Analysis of the HD 169142 Planet-Forming Disk}

\input{authors}

\begin{abstract}
We present a dynamical analysis of the HD 169142 planet-forming disk based on high-contrast polarimetric imaging over a twelve-year observational period, offering insights into its disk evolution and planet-disk interactions. This study explores the evolution of scattered-light features and their relationship with millimeter continuum emission. Archival visible-to-near-infrared scattered-light observations from NACO, SPHERE, and GPI combined with new observations from SCExAO reveal persistent non-axisymmetric structures in both the inner and outer rings of the disk. Through Keplerian image transformations and phase cross-correlation techniques, we show that the azimuthal brightness variations in the inner ring follow the local Keplerian velocity, suggesting these are intrinsic disk features rather than planet-induced spirals or shadows. The motion of the outer ring is weakly detected, requiring a longer observational baseline for further confirmation. Comparing scattered-light features with ALMA \SI{1.3}{\milli\meter}-continuum data, we find that the scattered light traces the edges of dust structures in the inner ring, indicating complex interactions and a leaky dust trap around the water-ice snowline. These findings highlight the capability of long-term monitoring of circumstellar disks to distinguish planetary influences from Keplerian disk dynamics.
\end{abstract}

\input{01_introduction}

\input{02_HD169142}
\input{03_observations}

\input{04_results}

\input{05_dynamic_analysis}

\input{06_compare_to_alma}
\input{07_discussions}
\input{08_conclusions}
\input{acknowledgements}

\facility{Subaru (SCExAO)}

\software{
    \texttt{ADEPTS} \citep{tobin_automated_2020,tobin_status_2022},
    \texttt{astropy} \citep{astropy_collaboration_astropy_2013,astropy_collaboration_astropy_2018},
    \texttt{CHARIS-DEP} \citep{brandt_data_2017},
    \texttt{CHARIS-DPP} \citep{currie_-sky_2020,lawson_high-contrast_2021},
    \texttt{diskmap} \citep{stolker_scattered_2016},
    Julia \citep{bezanson_julia_2017},
    \texttt{matplotlib} \citep{hunter_matplotlib_2007},
    \texttt{numpy} \citep{harris_array_2020},
    \texttt{pandas} \citep{team_pandas-devpandas_2024},
    \texttt{proplot} \citep{davis_proplot_2021},
    Python,
    \texttt{scikit-image} \citep{walt_scikit-image_2014},
    \texttt{showyourwork} \citep{luger_mapping_2021},
    \texttt{vampires-dpp}
}

\bibliography{references}

\appendix
\input{appendix_01_crosscors}

\input{appendix_02_data}

\end{CJK*}
\end{document}

%% file: authors.tex
\author[0000-0001-6341-310X]{Miles Lucas}
\affiliation{Steward Observatory, University of Arizona, Tucson, AZ 87521, USA}
\affiliation{Subaru Telescope, National Astronomical Observatory of Japan, 650 N. Aohoku Pl., Hilo, HI 96720, USA}
\email{mileslucas@hawaii.edu}

\author[0000-0003-1341-5531]{Michael Bottom}
\affiliation{Institute for Astronomy, University of Hawai'i, 640 N. Aohoku Pl., Hilo, HI 96720, USA}
\email{mbottom@hawaii.edu}

\author[0000-0001-9290-7846]{Ruobing Dong (董若冰)}
\affiliation{Kavli Institute for Astronomy and Astrophysics, Peking University, Beijing 100871, China}
\affiliation{Department of Physics and Astronomy, University of Victoria, Victoria, BC, V8P 5C2, Canada}
\email{rbdong@pku.edu.cn}

\author[0000-0002-7695-7605]{Myriam Benisty}
\affiliation{Max-Planck-Institut f\"ur Astronomie, K\"onigstuhl 17, D-67117 Heidelberg, Germany}
\email{benisty@mpia.de}

\author[0000-0002-9298-3029]{Mario Flock}
\affiliation{Max-Planck-Institut f\"ur Astronomie, K\"onigstuhl 17, D-67117 Heidelberg, Germany}
\email{flock@mpia.de}

\author[0000-0001-5763-378X]{Maria Vincent}
\affiliation{Institute for Astronomy, University of Hawai'i, 640 N. Aohoku Pl., Hilo, HI 96720, USA}
\email{mariavin@hawaii.edu}

\author[0000-0001-5058-695X]{Jonathan Williams}
\affiliation{Institute for Astronomy, University of Hawai'i, 2680 Woodlawn Dr., Honolulu, HI 96826, USA}
\email{jw@hawaii.edu}

\author[0000-0002-1094-852X]{Kyohoon Ahn}
\affiliation{Subaru Telescope, National Astronomical Observatory of Japan, 650 N. Aohoku Pl., Hilo, HI 96720, USA}
\affiliation{Korea Astronomy and Space Science Institute, 776 Daedeok-daero, Yuseong-gu, Daejeon 34055, Republic of Korea}
\email{kyohoon@naoj.org}

\author[0000-0002-7405-3119]{Thayne Currie}
\affiliation{Subaru Telescope, National Astronomical Observatory of Japan, 650 N. Aohoku Pl., Hilo, HI 96720, USA}
\affiliation{Department of Physics and Astronomy, University of Texas at San Antonio, One UTSA Circle, San Antonio, TX 78249, USA}
\email{Thayne.Currie@utsa.edu}

\author[0000-0003-4514-7906]{Vincent Deo}
\affiliation{Optical Sharpeners SAS, Manosque, France}
\affiliation{Subaru Telescope, National Astronomical Observatory of Japan, 650 N. Aohoku Pl., Hilo, HI 96720, USA}
\email{vdeo@naoj.org}

\author[0000-0002-1097-9908]{Olivier Guyon}
\affiliation{Subaru Telescope, National Astronomical Observatory of Japan, 650 N. Aohoku Pl., Hilo, HI 96720, USA}
\affiliation{Steward Observatory, University of Arizona, Tucson, AZ 87521, USA}
\affiliation{College of Optical Sciences, University of Arizona, Tucson, AZ 87521, USA}
\affiliation{Astrobiology Center, 2 Chome-21-1, Osawa, Mitaka, Tokyo, 181-8588, Japan}
\email{guyon@naoj.org}

\author[0000-0002-9294-1793]{Tomoyuki Kudo}
\affiliation{Subaru Telescope, National Astronomical Observatory of Japan, 650 N. Aohoku Pl., Hilo, HI 96720, USA}
\email{kudotm@naoj.org}

\author[0009-0009-6274-6514]{Lucinda Lilley}
\affiliation{Sydney Institute for Astronomy, School of Physics, Physics Road, University of Sydney, NSW 2006, Australia}
\email{lucinda.lilley@sydney.edu.au}

\author[0000-0002-3047-1845]{Julien Lozi}
\affiliation{Subaru Telescope, National Astronomical Observatory of Japan, 650 N. Aohoku Pl., Hilo, HI 96720, USA}
\email{lozi@naoj.org}

\author[0000-0001-6205-9233]{Maxwell Millar-Blanchaer}
\affiliation{Department of Physics, University of California, Santa Barbara, CA, 93106, USA}
\email{maxmb@ucsb.edu}

\author[0000-0002-8352-7515]{Barnaby Norris}
\affiliation{Sydney Institute for Astronomy, School of Physics, Physics Road, University of Sydney, NSW 2006, Australia}
\affiliation{Sydney Astrophotonic Instrumentation Laboratories, Physics Road, University of Sydney, NSW 2006, Australia}
\affiliation{AAO-USyd, School of Physics, University of Sydney, NSW 2006, Australia}
\email{barnaby.norris@sydney.edu.au}

\author[0000-0003-2953-755X]{Sebasti\'an P\'erez}
\affiliation{Departamento de F\'isica, Universidad de Santiago de Chile, Av. Victor Jara 3659, Santiago, Chile}
\affiliation{Millennium Nucleus on Young Exoplanets and their Moons, Chile}
\affiliation{Center for Interdisciplinary Research in Astrophysics and Space Science, Universidad de Santiago de Chile}
\email{sebastian.perez.ma@usach.cl}

\author[0000-0003-1713-3208]{Boris Safonov}
\affiliation{Sternberg Astronomical Institute, Lomonosov Moscow State Univeristy, 119992 Universitetskii prospekt 13, Moscow, Russia}
\email{safonov@sai.msu.ru}

\author[0000-0001-7026-6291]{Peter Tuthill}
\affiliation{Sydney Institute for Astronomy, School of Physics, Physics Road, University of Sydney, NSW 2006, Australia}
\email{peter.tuthill@sydney.edu.au}

\author[0000-0002-6879-3030]{Taichi Uyama}
\affiliation{Department of Physics and Astronomy, California State University, Northridge, Northridge, CA 91330 USA}
\email{taichi.uyama.astro@gmail.com}

\author[0000-0003-4018-2569]{S\'ebastien Vievard}
\affiliation{Space Science and Engineering Initiative, College of Engineering, University of Hawaii at Manoa, 640 N. Aohoku Pl., Hilo, HI 96720, USA}
\affiliation{Subaru Telescope, National Astronomical Observatory of Japan, 650 N. Aohoku Pl., Hilo, HI 96720, USA}
\email{vievard@naoj.org}

\author[0000-0003-3567-6839]{Manxuan Zhang}
\affiliation{Department of Physics, University of California, Santa Barbara, CA, 93106, USA}
\email{manxuanzhang@ucsb.edu}

%% file: 01_introduction.tex
\section{Introduction\label{sec:introduction}}

The diversity observed in exoplanetary systems may stem from the varying physical properties of protoplanetary disks, which set the initial conditions for planet formation \citep{williams_protoplanetary_2011}. As planets form and evolve within these disks, their interactions with the disk can significantly alter its structure, with massive planets exerting the most influence \citep{goodman_planetary_2001,wolf_observability_2005}. Understanding the concurrent processes of planet formation and disk evolution requires detailed knowledge of disk structures and planet-disk interactions. Direct imaging of gas and dust in protoplanetary disks provides essential insights into these processes.

Direct imaging of protoplanetary disks demands both high angular resolution and sensitivity. The characteristic scale of disk substructures is set by the local pressure scale height, typically on the order of $\sim$\SI{10}{\au} at a radius of $\sim$\SI{100}{\au} \citep{andrews_observations_2020}. At the distances of nearby star-forming regions ($d \sim$\SI{150}{\parsec}), resolving such features requires angular resolutions better than \SI{70}{\mas}. Thermal emission from cold millimeter-sized dust grains and pebbles is detectable at radio wavelengths where the stellar contribution is minimal, resulting in high disk-to-star contrast \citep{williams_protoplanetary_2011}. The Atacama Large Millimeter/sub-millimeter Array (ALMA) has leveraged these advantages to uncover a diversity of disk substructures that were previously inaccessible due to limitations in angular resolution and sensitivity \citep[e.g.,][]{andrews_disk_2018,perez_disk_2018,pinte_nine_2020}.

Scattered-light imaging at visible to near-infrared wavelengths offers a complementary view, tracing starlight reflected by sub-micron-sized dust grains in the disk surface layers \citep{benisty_optical_2023}. These grains have a small Stokes number, which means they remain dynamically coupled to the gas in the disk and are suspended at higher altitudes above the mid-plane compared to the larger grains probed by millimeter emission \citep{weidenschilling_aerodynamics_1977}. By leveraging observations across multiple wavelengths, it becomes possible to spatially resolve distinct vertical layers and compositional components within the disk, enabling a more comprehensive, three-dimensional characterization of disk morphology and evolution. However, scattered-light imaging presents additional challenges: the required angular resolution demands large telescope apertures with adaptive optics correction \citep{guyon_extreme_2018}, and the disk signal is typically three or more orders of magnitude fainter than the central star \citep{benisty_optical_2023,ginski_sphere_2024}. These constraints necessitate the use of high-contrast imaging systems capable of both resolving fine disk structures and suppressing stellar diffraction to isolate the faint circumstellar signal.

High-contrast polarimetry has emerged as a particularly effective technique for overcoming these challenges, exploiting the fact that starlight scattered by circumstellar dust is partially polarized--typically between 20\% and 70\% at peak \citep{tazaki_effect_2019,ren_karhunen-loeve_2023}--while the direct stellar light remains unpolarized. By splitting and subtracting orthogonal polarization states, polarimetric differential imaging (PDI) suppresses the unpolarized stellar component, isolating the polarized scattered light from the disk \citep{kuhn_imaging_2001}. Compared to traditional point-spread function (PSF) subtraction techniques, which are prone to self-subtraction artifacts and morphological biases \citep{lafreniere_new_2007, soummer_detection_2012}, PDI is less susceptible to residual speckle noise and yields higher-fidelity detections of faint disk structures in close proximity to bright stars.

Surveys have already detected most bright disks with polarimetry in nearby star-forming regions (distances of \SIrange{150}{400}{\parsec}), limited by the wavefront correction at visible wavelengths ($m_{\text R}<$12) (e.g., \citealt{avenhaus_disks_2018,garufi_disks_2020,rich_gemini-lights_2022,ren_protoplanetary_2023,valegard_sphere_2024,ginski_sphere_2024,garufi_sphere_2024}). The next frontier in disk science involves moving beyond detection to detailed characterization, which requires both multi-wavelength and multi-epoch observations to probe the dynamical processes shaping these systems.

Dynamical analysis of protoplanetary disks is crucial for understanding planet formation, disk evolution, and the interactions between gas, dust, and forming planets. Embedded planets sculpt disks, generating spiral arms, cavities, warps, and shadows, which move at different speeds than the local Keplerian disk velocity  \citep{goodman_planetary_2001,wolf_observability_2005,dong_what_2017,dong_how_2017,andrews_observations_2020}. Studying how these structures evolve provides indirect evidence of forming planets and their interactions with surrounding material. However, few dynamical studies have been conducted \citep[e.g.,][]{boccaletti_fast-moving_2015,debes_chasing_2017,stolker_variable_2017,ren_decade_2018,bertrang_moving_2020,sallum_systematic_2023,skaf__2023,avsar_search_2024}, and expanding these studies to more disks is key for analyzing substructures and their connection to planet formation.

In the case of face-on or nearly face-on disks, azimuthal features move with Keplerian velocity for their given stellocentric distance, $r$, and host star mass, $M_\star$--
\begin{equation}
\label{eqn:kep_third_law}
    \omega = \sqrt{\frac{GM_\star}{r^3}},
\end{equation}
where $G$ is the gravitational constant. This motion traces an arc with a length 
\begin{equation}
\label{eqn:arc_length}
    L = r\cdot\omega =\sqrt{\frac{GM_\star}{r}},
\end{equation}
which is resolved when the total displacement is at least one diffraction element. The distance subtended by one diffraction element depends on the observing wavelength, $\lambda$, aperture diameter, $D$,  and distance, $d$,
\begin{equation}
\label{eqn:diffraction}
    l =\frac{\lambda}{D}\cdot d.
\end{equation}
Dividing \cref{eqn:diffraction} by \cref{eqn:arc_length} creates a single expression for the time required to resolve Keplerian motion--
\begin{equation}
\label{eqn:resolved_motion_period}
    T= 0.035~\mathrm{yr}\cdot
    \left( \frac{\lambda}{\mathrm{\mu m}}\right)  
    \left( \frac{D}{\mathrm{m}}\right)^{-1}
    \left( \frac{d}{\mathrm{pc}}\right)
    \left(\frac{r}{\mathrm{au}} \right)^{1/2}
    \left(\frac{M_\star}{\mathrm{M_\odot}} \right)^{-1/2}
    .
\end{equation}
For example, using an \SI{8}{\meter} telescope at H-band ($\lambda=$\SI{1.6}{\micro\meter}) to observe a sun-like system at \SI{150}{\parsec} requires $\sim$\SI{11}{\year} to resolve the azimuthal motion of a feature \SI{100}{\au} from the star (roughly the extent of protoplanetary disks). Closer to the star, orbital velocities are higher, and shorter baselines suffice. For example, features at \SI{15}{\au} (close to the inner working angle of a coronagraphic mask projected at \SI{150}{\parsec}) are resolved after only \SI{4}{\year}. More massive stars, like Herbig Ae/Be systems \citep{brittain_herbig_2023}, will create faster orbital velocities, reducing the observational cadence needed to detect motion. 

Practically, shifts smaller than one diffraction element can be measured. For example, fitting a Gaussian or quadratic profile provides a centroid uncertainty that scales directly with feature width and inversely with the signal-to-noise ratio,
meaning that small, bright features can be tracked with high precision \citep[e.g.][]{ren_dynamical_2020}. Therefore, the temporal baselines required for meaningful dynamical analysis are shorter than those for resolving motion. As a result, multi-epoch imaging of nearly face-on disks on timescales of only a few years provides a viable path toward tracking Keplerian rotation and identifying deviations from it. 
Such dynamical studies will become increasingly powerful and widely applicable as more high-resolution scattered-light images are published and archived.

In this paper, we present a dynamical analysis of the planet-forming disk HD 169142, leveraging a twelve-year baseline of polarimetric images, combining archival and new observations to study the motion of non-axisymmetric disk features in scattered light.

%% file: 02_HD169142.tex
\section{The HD 169142 system\label{sec:hd169142}}

HD 169142 is a \SI{7.4\pm3.2}{\mega\year}-old \citep{blondel_modeling_2006,grady_disk_2007,rich_gemini-lights_2022} Herbig Ae star located at a distance of \SI{114.87\pm0.35}{\parsec} \citep{gaia_collaboration_gaia_2023}, with an estimated mass of \SI{1.89\pm0.08}{\Msun} \citep{blondel_modeling_2006,grady_disk_2007,rich_gemini-lights_2022,bohn_probing_2022}. It hosts a nearly face-on protoplanetary disk ($i=$\SI{12.58\pm0.48}{\degree}, $PA=$\SI{5.87\pm0.55}{\degree}; \citealp{raman_keplerian_2006,perez_dust_2019,bohn_probing_2022}) that exhibits multiple ring-like structures, potentially indicative of ongoing planetary formation. The fiducial system parameters were compiled from multiple sources, using an inverse-variance-weighted average to estimate parameters from multiple published values (\cref{tbl:system}).

\begin{deluxetable}{lcl}
    \centering
    \tablefontsize{\scriptsize}
    \tablecaption{Adopted system parameters for HD 169142.\label{tbl:system}}
    \tablehead{
        \colhead{Parameter} &
        \colhead{Value} & 
        \colhead{Ref.}
    }
    \startdata
    \cutinhead{System parameters}
    Distance (\si{pc}) & \num{114.87\pm0.35} & [1] \\
    $M_\star$ ($M_\odot$) & \num{1.89\pm0.08} & [2-4]\\
    $t_\star$ (\si{\mega\year}) & \num{7.4\pm3.2} & [2,4,5]\\
    \cutinhead{Global disk parameters}
    Inclination (\si{\degree}) & \num{12.58\pm0.48} & [3,6] \\
    Position angle (\si{\degree}) & \num{5.87\pm0.55} & [3,6] \\
    $M_\text{disk} / M_\star$ & \numrange{3e-3}{3e-2} & [7,8] \\
    \enddata
    \tablerefs{[1]: \citealt{gaia_collaboration_gaia_2023}, [2]: \citealt{blondel_modeling_2006}, [3]: \citealt{bohn_probing_2022}, [4]:  \citealt{rich_gemini-lights_2022}, [5]: \citealt{grady_disk_2007}, [6]: \citealt{raman_keplerian_2006} [7]: \citealt{fedele_alma_2017}, [8]: \citealt{perez_dust_2019}}
\end{deluxetable}

Spectroscopic studies of HD 169142 have detected strong infrared excess, polycyclic aromatic hydrocarbon (PAH) emission features, and variable near-infrared flux \citep{meeus_iso_2001,boekel_10_2005,meeus_gas_2010}. The mid-infrared spectrum lacks a prominent \SI{10}{\micro\meter} silicate feature, implying a depletion of small dust grains in the inner disk, possibly due to grain growth or clearing by forming planets.

\begin{figure}
    \centering
    \script{plot_disk_schematic.py}
    \includegraphics[width=\columnwidth]{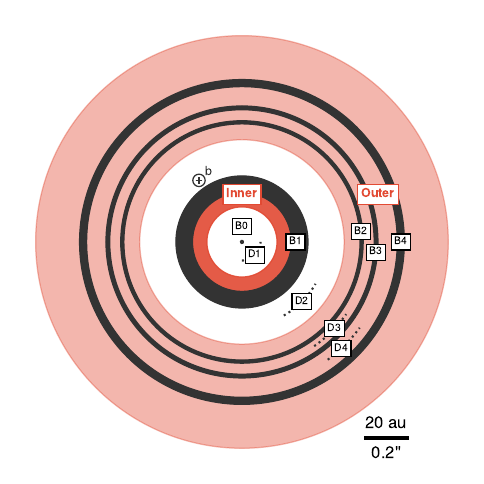}
    \caption{Schematic diagram of the HD169142 disk. The black contours are based on the ALMA \SI{1.3}{\milli\meter} continuum data which trace out the warped inner envelope (B0), inner cavity (D1), a close inner ring (B1), a ring gap (D2), and three outer rings (B2-B4) with intermediate gaps (D3, D4). The red contours are based on visible to near-IR scattered-light images, separated into an inner and outer ring. The location of the candidate protoplanet HD 169142 b from \citet{hammond_confirmation_2023} is marked with a circled cross.\label{fig:disk_schematic}}
\end{figure}

HD 169142's disk has been studied from radio to visible wavelengths, revealing a complex morphology. A schematic of the disk with the naming conventions from \citet{perez_dust_2019,poblete_protoplanetary_2022} is shown in \cref{fig:disk_schematic}. The disk is comprised of a warped (misinclined) circumstellar dust envelope (B0; \citealp{bohn_probing_2022}), a gas-depleted inner cavity within \SI{20}{\au} (D1), a bright ring at \SIrange{21}{26}{\au} (B1), a large ring gap at \SI{41}{\au} (D2), and an eccentric outer ring at \SI{66}{\au}. The outer ring contains three dust rings with two intermediate gaps in the \si{\milli\meter} dust emission (B2-B4, D3-D4; \citealp{perez_dust_2019}). 

High-contrast imaging, polarimetry, and millimeter observations of localized structures in the HD 169142 disk suggest that its features may be shaped by embedded protoplanets, especially the large annular ring gap \citep{honda_mid-infrared_2012,quanz_gaps_2013,reggiani_discovery_2014,biller_enigmatic_2014,fedele_alma_2017,macias_imaging_2017,ligi_investigation_2018,gratton_blobs_2019,macias_characterization_2019,perez_dust_2019,hammond_confirmation_2023}. Hydrodynamical models have predicted the presence of at least two planetary-mass companions, with the largest being $\sim$\SI{2}{\Mjup} \citep{fedele_alma_2017,pohl_circumstellar_2017,perez_dust_2019,toci_long-lived_2019,yu_mapping_2021,garg_kinematic_2022}. Most notably, a clump at a separation of \SI{37.2\pm1.5}{\au} in the large ring gap was detected across multiple epochs with Keplerian motion and reported as a protoplanet, HD 169142 b, initially in \citet{gratton_blobs_2019} and followed up by \citet{hammond_confirmation_2023}. In \citet{bertrang_hd_2018,bertrang_moving_2020}, the authors reported that a dip in the scattered-light intensity moved with super-Keplerian velocity, caused by shadowing from an inner companion at \SI{13}{\au}.

Previous studies using L$^{\prime}$-band coronagraphic imaging, which are sensitive to the thermal emission from giant planets, have reported the detection of a point source at \SI{22.7\pm4.7}{\au} \citep{reggiani_discovery_2014,biller_enigmatic_2014}. Photometric estimates suggested a massive object of approximately $\sim$\SI{30}{\Mjup}. However, follow-up near-infrared observations determined that the blob that had been detected was likely signal from the inner ring filtered by the PSF subtraction process \citep{ligi_investigation_2018}.

More recently, \citet{wallack_survey_2024} presented Keck/NIRC2 L$^{\prime}$-band observations that reach 5$\sigma$ contrast limits of \num{1.2e-3} at \ang{;;0.2} (\SI{23}{\au}) and \num{4e-4} at \ang{;;0.3} (\SI{37}{\au}). Using the ATMO-CEQ atmospheric and evolutionary models \citep{phillips_new_2020}, we converted these contrast limits into mass sensitivities assuming a system age of \SI{7.4\pm3.2}{\mega\year}. The inferred detection thresholds are \SIrange{2.3}{3.7}{\Mjup} at \SI{23}{\au} and \SIrange{1.4}{2.2}{\Mjup} at \SI{37}{\au}, in agreement with previous constraints in the literature \citep{reggiani_discovery_2014,biller_enigmatic_2014}. These non-detections place meaningful upper limits on potential companions, grounded in direct imaging rather than indirect inference.

Given the complex morphology and the presence of non-axisymmetric structures in the HD 169142 disk, dynamical analysis is useful for understanding the mechanisms shaping these features. The detection of a candidate protoplanet at \SI{37}{\au}, along with the identification of potential companions at \SI{23}{\au} and \SI{13}{\au}, further emphasizes the need for a detailed investigation into planet-disk planet interactions and the resulting dynamical evolution. By tracking the motion of disk features over time, such analysis can distinguish between local perturbations induced by embedded objects and intrinsic disk features, offering insights into the early stages of planet formation in this system.

%% file: 03_observations.tex
\section{Observations and Data Reduction\label{sec:observations}}

In this work, we present a combination of new high-contrast polarimetric observations of HD 169142 using the Subaru Coronagraphic Extreme Adaptive Optics instrument (SCExAO; \citealp{jovanovic_subaru_2015}) at the Subaru telescope alongside previously published and archival polarimetric datasets. All observations are listed in \cref{tbl:obslog}.

\begin{deluxetable*}{llllccccccl}
\tabletypesize{\footnotesize}
\tablehead{
    \colhead{Date} &
    \multirow{2}{*}{Telescope} &
    \multirow{2}{*}{Instrument} &
    \multirow{2}{*}{Filter} &
    \colhead{IWA} &
    \colhead{pix. scale} &
    \colhead{DIT} &
    \colhead{$T_\mathrm{exp}$} &
    \colhead{Seeing} &
    \multirow{2}{*}{Ref.} \vspace{-0.75em} \\
    \colhead{(UTC)} & 
    & 
    & 
    & 
    \colhead{(mas)} & 
    \colhead{(mas/pix)} & 
    \colhead{(\si{\second})} & 
    \colhead{(\si{min})} & 
    \colhead{('')} & 
     & 
}
\tablecaption{Chronologically-ordered observing log.\label{tbl:obslog}}
\startdata
\formatdate{26}{07}{2012} & VLT & NACO & H & sat. & 27 & 45 & 69 & $\sim$1.04 & [1] \\
\formatdate{25}{04}{2014} & Gemini-S & GPI & J & 184 & 14.14 & 29.1 & 62 & \num{0.85\pm0.27} & [2] \\
\formatdate{03}{05}{2015} & VLT & IRDIS & J & 80 & 12.263 & 16 & 53 & $\sim$0.9 & [3] \\ 
\formatdate{10}{07}{2015} & VLT & ZIMPOL & VBB & sat. & 3.6 & 10 & 56 & 0.75 & [4] \\
\formatdate{15}{07}{2018} & VLT & ZIMPOL & VBB & sat. & 3.6 & 4.4 & 53 & 0.51 & [5]  \\
\formatdate{06}{09}{2021} & VLT & IRDIS & Ks & 200 & 12.265 & 16 & 34 & $\sim$0.5 & [6] \\
\formatdate{3}{6}{2023} & Subaru & CHARIS & JHK & 113 & 15.16 & 60 & 52 & \num{0.82\pm0.15}$^a$ & \\ 
\formatdate{4}{6}{2023} & Subaru & CHARIS & JHK & 113 & 15.16 & 60 & 40 &  & \\ 
\formatdate{29}{7}{2024} & Subaru & VAMPIRES & MBI & 54 & 5.9 & 2 & 108 & \num{0.6\pm0.14}$^a$ & \\
\enddata
\tablecomments{IWA: inner working angle of coronagraph, if present, or ``sat.'' for saturated non-coronagraphic observations. DIT: detector integration time for each frame. $T_\text{exp}$ total exposure time. (a): Seeing estimates, where present, were retrieved from the mean and standard deviation DIMM from the Maunakea Weather Center archive (\url{http://mkwc.ifa.hawaii.edu/current/seeing/index.cgi}).}
\tablerefs{[1]: \citealt{quanz_gaps_2013}, [2]: \citealt{monnier_polarized_2017}, [3]: \citealt{pohl_circumstellar_2017}, [4]: \citealt{bertrang_hd_2018}, [5]: \citealt{bertrang_moving_2020}, [8]: \citealt{ren_protoplanetary_2023}}

\end{deluxetable*}
\subsection{SCExAO/CHARIS Observations\label{sec:obs_charis}}

We observed HD 169142 over two nights in 2023 (proposal ID: o23164) using the CHARIS integral-field spectrograph in broadband PDI mode \citep{groff_charis_2015,groff_first_2017}. The combination of a dispersing prism and micro-lens array creates spectra across the field of view from J to K band (\SIrange{1.12}{2.4}{\micro\meter}) and a Wollaston prism splits the ordinary and extraordinary beams which are simultaneously imaged on the detector. A rectangular \ang{;;1}x\ang{;;2} field stop prevents the two orthogonal beams from overlapping on the micro-lens array, which creates a distinctive wedge shape in derotated and collapsed data (e.g., \cref{fig:qphi_mosaic}). An upstream half-wave plate (HWP) was modulated between \SIlist{0;45;22.5;67.5}{\degree} to measure the linear Stokes parameters ($Q$ and $U$) and to correct for instrumental polarization. The \SI{113}{mas} Lyot coronagraph was used with calibration ``satellite spot'' speckles for alignment and photometric calibration \citep{sahoo_precision_2020}.

We extracted calibrated three-dimensional spectral cubes from each night's data using the \texttt{ADEPTS} pipeline \citep{brandt_data_2017,tobin_automated_2020,tobin_status_2022}. Then, we used \texttt{CHARIS-DPP} \citep{currie_-sky_2020,lawson_high-contrast_2021} for image registration and spectrophotometric calibration using a template stellar spectrum \citep{castelli_new_2004}. For polarimetry, we derotated each data cube and applied an inverse-least squares Mueller-matrix correction \citep{joost_t_hart_full_2021}, which produced calibrated Stokes Q and U frames for each wavelength slice. The stellar polarization was measured within the coronagraph mask and removed from the Stokes Q and U cubes. We combined the individual Stokes Q and U cubes from both nights, discarding a small fraction (5\%) of frames with poor polarimetric signal-to-noise ratios (S/N). Finally, we collapsed the resulting Q and U spectral cubes with an outlier-resilient mean, excluding slices corresponding to the telluric absorption bands. We calculated the azimuthal Stokes images $Q_\phi$ and $U_\phi$ \citep{schmid_limb_2006,monnier_multiple_2019},
\begin{align}
\begin{split}
    \label{eqn:az_stokes}
    Q_\phi &= -Q\cos{\left(2\phi\right)} - U\sin{\left(2\phi\right)} \\
    U_\phi &= -Q\sin{\left(2\phi\right)} + U\cos{\left(2\phi\right)},
\end{split}
\end{align}
where
\begin{equation}
    \phi = \arctan{\left( \frac{x_\star - x}{y - y_\star} \right)} + \phi_0,
\end{equation}
the angle east of north plus any calibration offsets.

\subsection{SCExAO/VAMPIRES Observations\label{sec:obs_vampires}}

We observed HD 169142 in 2024 (proposal ID: o24184) using SCExAO/VAMPIRES \citep{norris_vampires_2015,lucas_visible-light_2024-1}. We used the multiband imaging mode (MBI), which simultaneously captures four \SI{50}{\nano\meter}-bandpass images centered at \SI{610}{\nano\meter}, \SI{670}{\nano\meter}, \SI{720}{\nano\meter}, and \SI{760}{\nano\meter}. We used the slow detector readout mode for low read noise and an upstream HWP for modulation to correct for instrumental polarization. We used a \SI{54}{mas} Lyot coronagraph with calibration speckles for alignment and photometric calibration.

We processed the data using \texttt{vampires-dpp}\footnote{\url{https://github.com/scexao-org/vampires_dpp}}, subtracting a master dark frame from each exposure before co-registering with sub-pixel phase cross-correlation \citep{guizar-sicairos_efficient_2008}. We used the flux measured from the calibration speckles \citep{sahoo_precision_2020,lucas_visible-light_2024} and a template spectrum \citep{pickles_stellar_1998} for spectrophotometric calibration. We grouped the aligned and calibrated frames into sequences according to the HWP angle and discarded 25\% of frames with the worst Strehl ratio in each sequence. We collapsed each sequences with a median before applying a least-squares Mueller-matrix correction, assuming ideal coefficients in the Mueller matrix, to produce calibrated Stokes I, Q, and U frames for each wavelength. The stellar polarization was measured within the coronagraph mask and removed from the Stokes cubes. Due to the lack of a calibrated Mueller matrix, the polarimetric images from VAMPIRES are not as accurate, however, the inner ring is still detected, and we avoid any quantitative polarimetry using this data. We collapsed the four MBI wavelengths before calculating the $Q_\phi$ and $U_\phi$ images.

\subsection{Archival Data\label{sec:obs_archive}}

We analyzed six archival polarimetric datasets of HD 169142 spanning from 2012 to 2021 for this study. The oldest dataset is the \formatdate{26}{7}{2012} H-band polarimetric observation from the Very Large Telescope (VLT) Nasmyth Adaptive Optics System Near-Infrared Imager and Spectrograph (NACO; \citealp{quanz_very_2011}) presented in \citet{quanz_gaps_2013}. The observations did not use a coronagraph, causing saturation around the central star which is masked out in the images. We used the stellar-polarization-corrected data presented in \citet{regt_polarimetric_2024} from the \texttt{PIPPIN} pipeline, which uses modern polarimetric data processing techniques in a consistent, purpose-built package for NACO.

The Gemini Planet Imager (GPI; \citealt{macintosh_first_2014}) observed HD 169142 on \formatdate{25}{04}{2014} in J-band non-dispersed polarimetric mode \citep{perrin_polarimetry_2015}. A \SI{92}{\mas} IWA apodized Lyot coronagraph (APLC) was used for diffraction control. The data was originally presented in \citet{monnier_polarized_2017}, and we use the data products from \citet{rich_gemini-lights_2022}, which have removed the stellar polarization.

The Spectro-Polarimetic High contrast imager for Exoplanets REsearch (SPHERE; \citealp{beuzit_sphere_2019}) has observed HD 169142 with the InfraRed Dual Imager and Spectrograph (IRDIS; \citealp{dohlen_infra-red_2008}) and the Zurich IMaging POLarimeter (ZIMPOL; \citealp{schmid_spherezimpol_2018}). On \formatdate{03}{05}{2015} it was observed with IRDIS in J-band polarimetric mode \citep{boer_polarimetric_2020} with an \SI{80}{\mas} IWA APLC, first presented in \citet{pohl_circumstellar_2017}. The data was processed with \texttt{IRDAP} \citep{holstein_polarimetric_2020}, using the calibrated Mueller-matrix correction and removing the stellar polarization fraction. HD 169142 was also observed on \formatdate{06}{09}{2021} in Ks-band polarimetric mode with a \SI{200}{\mas} IWA APLC, first presented in \citet{ren_protoplanetary_2023}. This data was reduced with Mueller-matrix correction and stellar polarization removal.

The ZIMPOL data were observed on \formatdate{10}{7}{2015} and \formatdate{15}{7}{2018} in the very broadband (VBB) filter, which spans \SIrange{590}{880}{\nano\meter}, in ``fastPol'' mode \citep{schmid_spherezimpol_2018}. The central PSF is saturated in these images to enable high dynamic range images without a coronagraph. The 2015 epoch was first presented in \citet{bertrang_hd_2018}, and the 2018 epoch was first presented in \citet{bertrang_moving_2020}. These data have not used a Mueller-matrix for correction, nor have they had the stellar polarization removed.

\begin{figure}
    \centering
    \script{plot_ALMA.py}
    \includegraphics[width=\columnwidth]{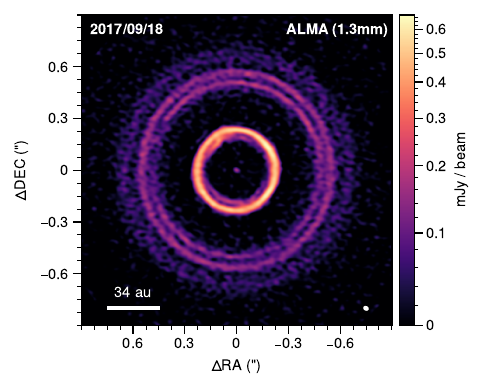}
    \caption{ALMA \SI{1.3}{\milli\meter} continuum observations from \citet{perez_disk_2018}. The data is shown with a slight asinh stretch. North is up, and east is left. The synthesized beam profile is shown with an ellipse in the bottom right.\label{fig:alma}}
\end{figure}

Lastly, we obtained the high-resolution ALMA \SI{1.3}{\milli\meter} continuum images observed on \formatdate{18}{9}{2017} and first presented in \citet{perez_dust_2019}. We used deconvolved images using the maximum entropy method (MEM; \citealt{carcamo_multi-gpu_2018}), with a synthesized beam size of \SI{27}{\mas}$\times$\SI{20}{\mas} and an RMS noise of \SI{14.7}{\micro Jy/beam}. The ALMA data is shown in \cref{fig:alma}.

%% file: 04_results.tex
\section{Results\label{sec:results}}
\subsection{Polarimetric Images\label{sec:pol_images}}

We used the azimuthal Stokes $Q_\phi$ images as the \textit{de facto} measurement of polarized intensity and the $U_\phi$ images as error maps, following the standard approach for analyzing scattered-light disk data \citep{schmid_limb_2006,avenhaus_disks_2018}. For centrosymmetric scattering geometries, such as those expected from protoplanetary disks illuminated by a central star, the linearly polarized light is predominantly oriented tangent to circles around the central source. In this geometry, $Q_\phi$ captures the azimuthal polarization component and thus contains nearly all of the astrophysical polarized signal, while $U_\phi$ ideally contains no disk signal and serves as a measure of noise or residual systematics.

\begin{figure*}
    \centering
    \script{plot_mosaics_Qphi.py}
    \includegraphics[width=\textwidth]{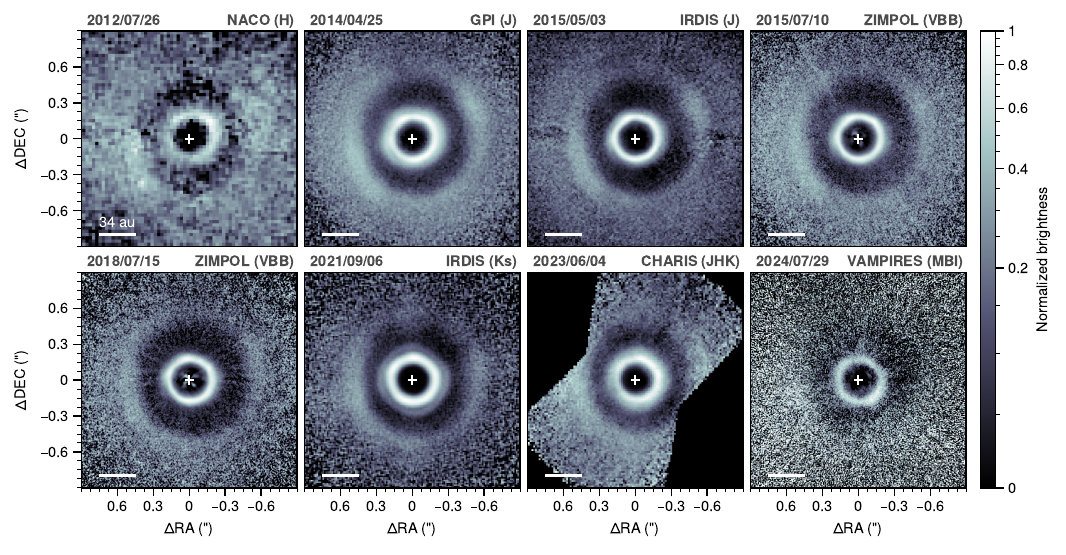}
    \caption{Polarized surface brightness images ($Q_\phi \times r^2$), shown with an asinh color scale to emphasize the cavity and outer ring. Each image is scaled from 0 to the maximum value in the inner ring (\SIrange{15}{35}{\au}). North is up and east is left in all images.\label{fig:qphi_mosaic}}
\end{figure*}

\begin{figure*}
    \centering
    \script{plot_mosaic_Uphi.py}
    \includegraphics[width=\textwidth]{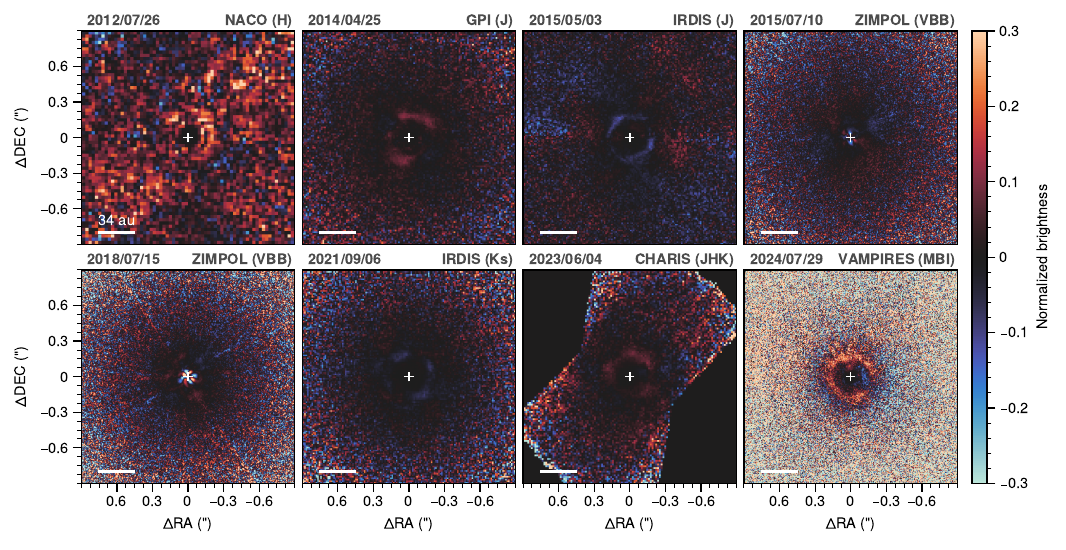}
    \caption{$U_\phi \times r^2$ images from each epoch, normalized to the same limits as \cref{fig:qphi_mosaic} and shown with a linear scale. A \SI{34}{\au} scale bar is shown in the bottom left of all panels. North is up and east is left in all images. \label{fig:uphi_mosaic}}
\end{figure*}

To quantify the polarized surface brightness, the polarimetric images must be corrected for the radial dependence of stellar irradiation, which decreases with the square of the stellocentric distance. We used the \texttt{diskmap} package \citep{stolker_scattered_2016} to compute the irradiation correction maps ($r^2$, in \si{au^2}), accounting for disk inclination and position angle. We used a flat disk model, neglecting effects of disk flaring, which might affect the brightness scaling of features far from the star, but which we do not expect to severely affect our analysis thanks to the low inclination of the disk. The resulting irradiation-corrected polarized surface-brightness images ($Q_\phi \times r^2$) are shown in \cref{fig:qphi_mosaic}. The corresponding $U_\phi \times r^2$ images are shown in \cref{fig:uphi_mosaic}.

\begin{figure*}
    \centering
    \script{plot_mosaics_Qphi_inner.py}
    \includegraphics[width=\textwidth]{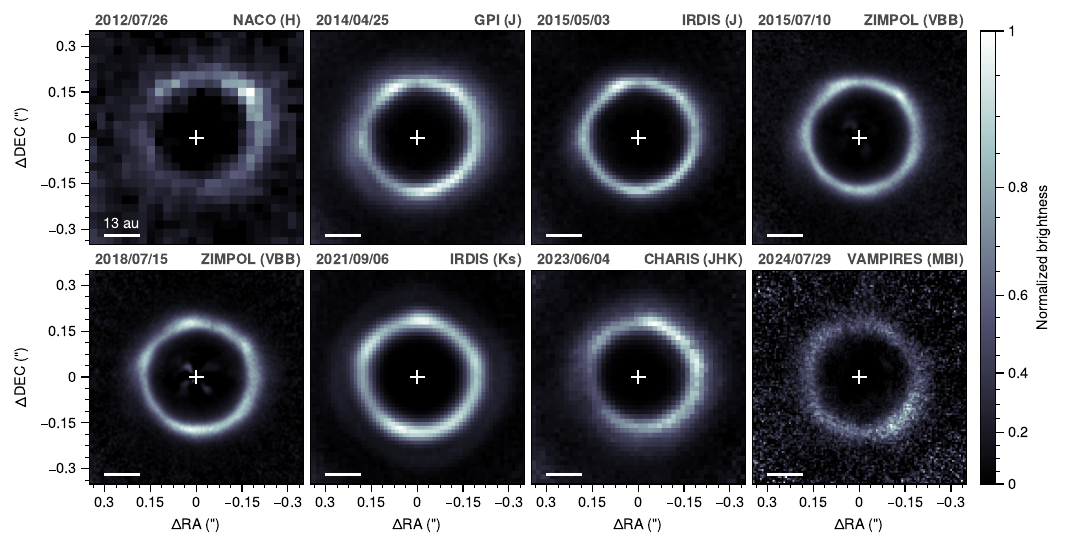}
    \caption{The same as \cref{fig:qphi_mosaic}, but zoomed into the inner ring, and shown with a sinh color scale.\label{fig:qphi_mosaic_inner}}
\end{figure*}

A bright, well-defined inner ring is visible in all observations, indicating strong scattering from small dust grains in the surface layers. This ring appears relatively axisymmetric but exhibits variations in intensity along its azimuth. Near-infrared observations (GPI J-band, IRDIS J- and Ks-band, CHARIS JHK) show a faint outer component of this inner ring, especially the IRDIS Ks-band image. While the ZIMPOL images show some wispy features in the same region, they are not nearly as bright as seen in the near-infrared. A magnified view of the inner ring is shown in \cref{fig:qphi_mosaic_inner}. A large gap between the inner and outer rings is consistently observed in all datasets. The outer ring has more pronounced azimuthal variations, with broad and diffuse peaks at $\sim$\ang{135} and $\sim$\ang{315}. 

In \cref{fig:uphi_mosaic}, we show the $U_\phi\times r^2$ images normalized to the same limits as the $Q_\phi\times r^2$ data. The NACO and VAMPIRES datasets show more noise in general than any other dataset. The non-coronagraphic ZIMPOL datasets have a unique octopole pattern (also present in the $Q_\phi$ data), which can be attributed to unresolved polarization (potentially from the inner disk, B0; \citealt{bertrang_hd_2018,tschudi_quantitative_2021}). The GPI, IRDIS, and ZIMPOL datasets all show structure within the inner ring. These features are a small fraction of the $Q_\phi$ signal in the same region, around 10\% or less. Significant $U_\phi$ signal can be attributed to multiple scattering; however, the lack of a consistent $U_\phi$ signal across instruments and epochs makes it more likely to be instrumental noise (e.g., polarimetric cross-talk) than astrophysical. 

\begin{figure*}
    \centering
    \script{plot_mosaic_Uphi.py}
    \includegraphics[width=\textwidth]{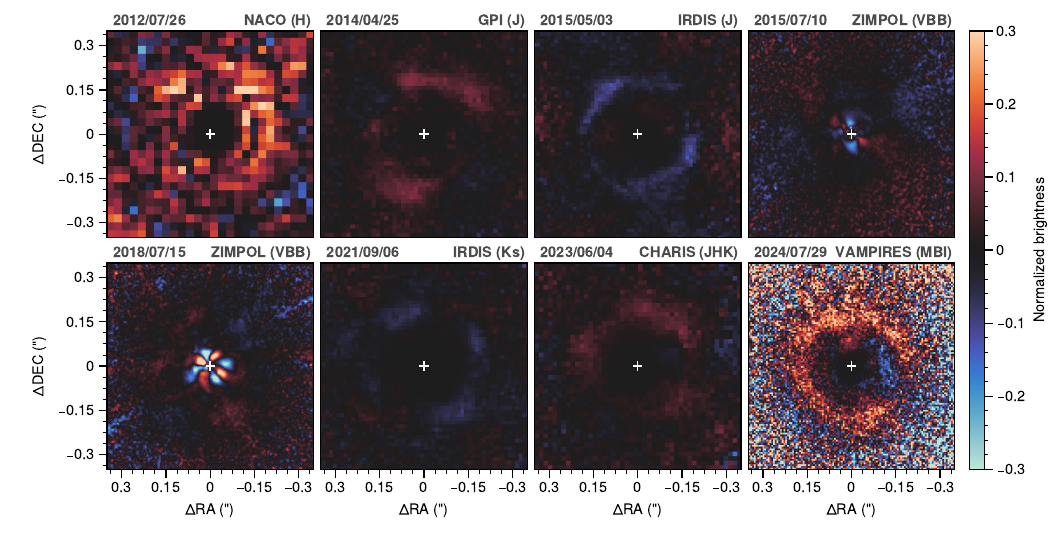}
    \caption{Similar to \cref{fig:uphi_mosaic}, but zoomed into the inner ring.\label{fig:uphi_mosaic_inner}}
\end{figure*}

\subsection{Radial Profiles\label{sec:radial_profiles}}

\begin{figure}
    \centering
    \script{plot_radial_profile_Qphi_combined_ALMA.py}
    \includegraphics[width=\columnwidth]{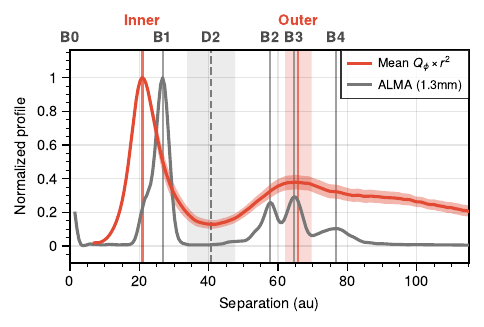}
    \caption{Radial profiles from the mean deprojected scattered light image corrected for stellar irradiation ($Q_\phi \times r^2$; red curve) and the projected ALMA \SI{1.3}{\milli\meter} dust continuum emission from \citealt{perez_dust_2019} (gray curve). Both curves are shaded with their 1$\sigma$ measurement uncertainty from the standard error. The five dust peaks (B0-B4) are denoted with gray vertical lines. The inner and outer scattered light peaks are shaded in red according to the 1$\sigma$ uncertainties measured from bootstrap resampling. The gap in the scattered light image (D2) is denoted with a gray dashed line, and the gap width is shaded in gray.\label{fig:radial_profiles}}
\end{figure}

We extracted radial profiles from the deprojected $Q_\phi \times r^2$ images, using the $U_\phi \times r^2$ images as error maps. We convolved the data and error images with a 1 pixel FWHM Gaussian kernel before measuring the profiles with 1 pixel wide radial bins. The radial profile and error curves from each dataset were interpolated onto a common grid before mean-combining the profiles-- we excluded the CHARIS data due to the large wedges without signal in the outer ring (\cref{fig:qphi_mosaic}). The mean profile and the ALMA \SI{1.3}{\milli\meter} continuum profile are shown in \cref{fig:radial_profiles}.

We measured the locations of the peaks in both the dust emission and scattered-light from the radial profiles. We measured the location and maximum intensity for each peak, using bootstrap resampling to derive uncertainties. The measured values and associated 1$\sigma$ uncertainties are reported in \cref{tbl:ring_parameters}. Additionally, we characterized the gap between the inner and outer rings (D2) following the methodology from \citet{dong_what_2017} to estimate the gap's width and depth. This involved determining the radius of minimum brightness, $r_\mathrm{min}$, the effective gap width, $w_I$, and the gap depth or gap contrast, $\delta_I$, from the radial profile of the scattered light image. Our results agree with previous characterizations of the dust and scattered light rings \citep{dong_what_2017,perez_dust_2019,bertrang_hd_2018,tschudi_quantitative_2021,ligi_investigation_2018}.

The inner ring shows a separation of $\sim$\SI{6}{\au} between the scattered-light and the millimeter dust emission. The inner ring profile in scattered-light has different slopes for the interior and exterior edges. The outer ring scattered-light profile has a complicated shape with a steep inner slope, a relatively flat peak, and a shallow but inconsistent outer slope, with inflection points around \SI{75}{\au} and \SI{98}{\au}.

\begin{deluxetable}{lc}
    \centering
    \tablecaption{HD 169142 ring and gap parameters.\label{tbl:ring_parameters}}
    \tablehead{
        \colhead{Parameter} &
        \colhead{Value}
    }
    \startdata
    \multicolumn{2}{l}{ALMA \SI{1.3}{\milli\meter} Dust Emission} \\
    \hline
    B1 ring $r_\text{max}$ (\si{\au}) & \num{26.708\pm0.006} \\
    B2 ring $r_\text{max}$ (\si{\au}) & \num{57.74\pm0.10} \\
    B3 ring $r_\text{max}$ (\si{\au}) & \num{64.66 \pm 0.15} \\
    B4 ring $r_\text{max}$ (\si{\au}) & \num{76.66 \pm 0.56} \\
    \hline
    \multicolumn{2}{l}{Mean $Q_\phi \times r^2$ Surface Brightness} \\
    \hline
    Inner ring $r_\text{max}$ (\si{\au}) & \num{20.9\pm0.4} \\
    Outer ring $r_\text{max}$ (\si{\au}) & \num{65.9\pm3.8} \\
    D2 gap $r_\text{min}$ (\si{\au}) & \num{40.5\pm2.9} \\
    D2 gap $w_{I}$ (\si{\au}) & \num{13.6\pm4.1} \\
    D2 gap $\delta_{I}$ & \num{4.8\pm1.2} \\
    D2 gap \(M_p\) (\si{\Mjup})         & [\num{0.2\pm0.1}, \num{0.7\pm0.5}, \num{2.2\pm1.5}] \\                                        & for \(\alpha=[10^{-4},\,10^{-3},\,10^{-2}]\) \\
    \enddata
    \tablecomments{Uncertainties are 1$\sigma$ errors estimated with bootstrap resampling (\cref{sec:radial_profiles}). The gap parameters are derived from \citealt{dong_what_2017}-- $r_\text{min}$: the location of minimum flux, $w_I$: gap width, $\delta_I$ gap depth (contrast).}
\end{deluxetable}

\subsection{Polar Projections}

\begin{figure}
    \centering
    \script{plot_polar_collapsed_inner.py}
    \includegraphics[width=\columnwidth]{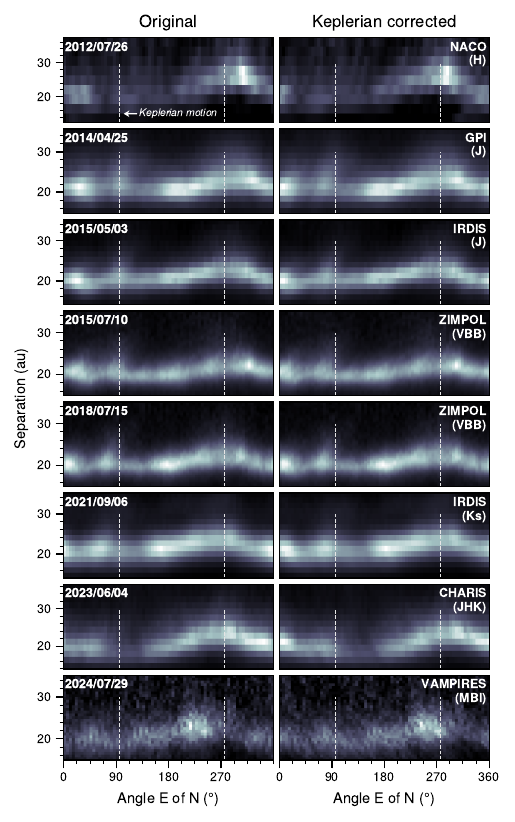}
    \caption{Polarized surface brightness ($Q_\phi \times r^2$) polar plot of the inner ring (\SIrange{15}{35}{au}). The left column is the original data, while the right column have been transformed to correct for Keplerian motion. The y-axis is separation in au. The observation date is in the top left while the instrument and filter combination is in the top right. The disk minor axis is marked with dashed white lines. A $\sinh$ stretch is used to emphasize features in the ring, and the values are normalized from 0 to the maximum value in the inner ring (the same as \cref{fig:qphi_mosaic_inner}). The direction of Keplerian rotation is shown with a white arrow in the top left plot.\label{fig:qphi_polar_inner}}
\end{figure}

\begin{figure}
    \centering
    \script{plot_polar_collapsed_outer.py}
    \includegraphics[width=\columnwidth]{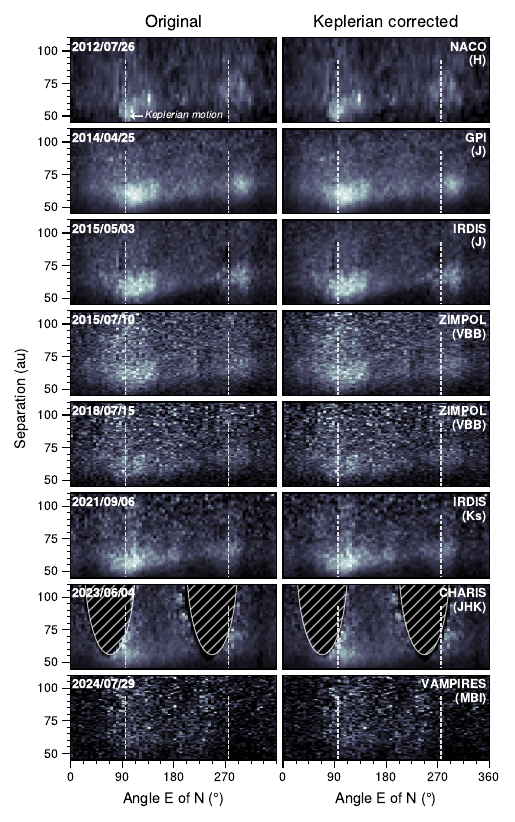}
    \caption{The same as \cref{fig:qphi_mosaic_inner}, but the disk's outer ring (\SIrange{45}{110}{\au}). The CHARIS data has missing wedges shown with cross-hatches due to the rectangular field stop.\label{fig:qphi_polar_outer}}
\end{figure}

We projected each dataset into polar coordinates to analyze the azimuthal and radial distributions of the inner and outer rings. We used \ang{5} wide azimuthal bins and 1 pixel wide radial bins to define the mapping from Cartesian to polar coordinates. The bin sizes were chosen to avoid sub-sampling our images during the interpolation onto the polar grid--ultimately limited by the angle subtended by one pixel at \SI{15}{\au}. We defined the inner ring region from \SIrange{15}{35}{\au}, and the outer ring region from \SIrange{48}{110}{\au}. The polar plots are shown in the left column of \cref{fig:qphi_polar_inner} and \cref{fig:qphi_polar_outer} for the inner and outer rings, respectively.

The polar projection of the inner ring shows consistent features across epochs. We refer to \cref{fig:qphi_polar_inner_annotated} for the annotations of the features. In the \SIrange{0}{90}{\degree} position angle range there are two bright clumps (C1 and C2) separated by $\sim$\ang{75}. There is an approximately \ang{30}-wide dip in between the clumps (S1), which is the same as the shadow reported in \citet{quanz_gaps_2013,bertrang_hd_2018,ligi_investigation_2018,bertrang_moving_2020}. There is another dip (S2) around \ang{50} wide to the right of the second clump, followed by the third bright clump (C3). Following this dip, the ring becomes brighter and wider and starts to bow out. This outward bowing could be explained by eccentricity, but that would create a sinusoidal profile with a contraction mirroring the bowing \ang{180} apart, which we do not see. \citet{perez_dust_2019} reached a similar conclusion, demonstrating that the millimeter dust emission of the inner ring is not well-fit by a circular or elliptical profile.

These non-axisymmetric features are moving over time. The direction of motion is from right to left in the polar plots, which is clockwise on the sky. This is consistent with the direction of Keplerian motion \citep{ligi_investigation_2018,garg_kinematic_2022,hammond_confirmation_2023}. 

The outer ring also has consistent scattered-light features across epochs and instruments, including bright clumps at \ang{90} and \ang{270}. The clump at \ang{90} is $\sim$\ang{75} wide, and the other clump is $\sim$\ang{55} wide. The outer ring does show an eccentric sinusoidal profile, which matches the analysis of the outer dust rings by \citet{perez_dust_2019}. It is more difficult to track the motion of the outer ring due to its distance from the star as well as the broader, more diffuse features compared to the inner ring. Following \cref{eqn:resolved_motion_period}, the outer disk motion is resolved within \SI{5}{\year} at \SI{1.6}{\micro\meter} on an \SI{8}{\meter} telescope, so the twelve-year baseline of our data is sufficient for meaningful analysis. Visually, when comparing the 2014 GPI epoch with the 2021 IRDIS epoch, there is motion of the bright features from right to left.

%% file: 05_dynamic_analysis.tex
\section{Dynamical Analysis of Disk Features\label{sec:dynamical_analysis}}

The observed motion--or lack of motion--of scattered-light features offers important constraints on the dynamical processes shaping protoplanetary disks, particularly those driven by interactions with embedded planets. Under the null hypothesis, disk material is assumed to orbit the star at the local Keplerian velocity, which is determined solely by the stellar mass and stellocentric distance. This assumes negligible self-gravitational effects in the disk, which is reasonable given the low disk-to-star mass ratio ($10^{\mathrm{-}3}$ to $10^{\mathrm{-}2}$; \citealt{fedele_alma_2017,perez_dust_2019}). While disks exhibit a slight non-Keplerian rotation due to radial pressure support in the gas, this deviation is negligible for the angular precision and timescales relevant to our analysis. Detecting significant departures from Keplerian motion would indicate the presence of additional dynamical influences--such as spiral density waves, vortices, shadowing, or disk warps. Therefore, measuring the angular velocities of scattered-light features provides a valuable diagnostic for probing the presence and influence of forming planets within the disk.

\subsection{Keplerian Image Shearing\label{sec:polar_warping}}

Our initial method for probing disk dynamics involves aligning scattered-light features across multiple epochs. This technique tests whether azimuthal motion is consistent with Keplerian rotation by transforming each image into a rotating reference frame defined by the expected Keplerian velocity. In this frame, structures that move with the local Keplerian speed will appear stationary over time, while any deviations--features that drift or blur--indicate sub- or super-Keplerian motion. By identifying which features remain aligned and which do not, we can begin to isolate dynamical signatures of planet-disk interactions or other processes that perturb the disk's orbital motion. 

To put our data into the Keplerian reference frame, we derive an image transformation based on a circular, Keplerian orbit in the disk midplane. The polar images in \cref{fig:qphi_polar_inner,fig:qphi_polar_outer} have surface brightnesses distributed over radius and azimuth, $\Sigma\left(r, \theta\right)$. The images are transformed by interpolating values at new coordinates, using an interpolation operator, $\mathcal{I}$,
\begin{equation}
    \Sigma\left(r', \theta' \right) = \mathcal{I}\left\{\Sigma\left(r, \theta\right)\right\}\left(r', \theta'\right).
\end{equation}
The new coordinates are defined using a time-dependent translation,
\begin{equation}
    \begin{bmatrix}
        r' \\ \theta'
    \end{bmatrix} =
    \begin{bmatrix}
        r \\ \theta
    \end{bmatrix} + \mathbf{b}\left(r, t - t_0\right),
\end{equation}
determined from \cref{eqn:kep_third_law}-- 
\begin{equation}
    \mathbf{b}\left(r, t - t_0\right) =
    \begin{bmatrix}
        0 \\
        \left(t - t_0\right)\cdot\omega(r)
    \end{bmatrix}.
\end{equation}
In other words, we are shearing the polar images along the azimuthal axis. The amount of shearing depends on the stellocentric distance and stellar mass, as well as the time elapsed since the observational epoch, $t_0$.

\begin{figure}
    \centering
    \script{plot_polar_combined_Qphi_inner.py}
    \includegraphics[width=\columnwidth]{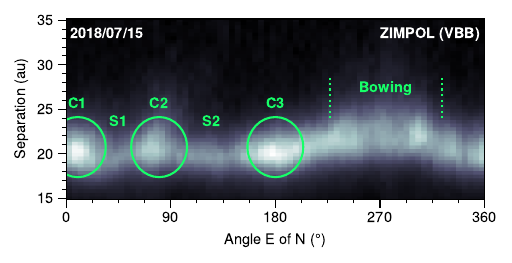}
    \caption{\formatdate{15}{7}{2018} ZIMPOL VBB $Q_\phi \times r^2$ polar image of the inner ring with annotations. C1-C3 are bright clumps, while S1 and S2 are dimmer shadow-like features. The region of outward bowing is marked with dotted lines.\label{fig:qphi_polar_inner_annotated}}
\end{figure}

We transformed the polar images in \cref{fig:qphi_polar_inner,fig:qphi_polar_outer} to a common time: \formatdate{15}{07}{2018}. This is roughly the midpoint of the observations, which minimizes the deshearing applied to the oldest and newest datasets. The Keplerian-corrected images are presented in the right column of the polar plots. The disk features highlighted in \autoref{sec:results} appear aligned in the inner ring images, including the bright clumps (C1-C3) from \ang{0} to \ang{180}, the shadow-like dips (S1 and S2) at \ang{45} and \ang{120}, and the bright, outward-bowing feature around $\sim$\ang{270}. The outer ring has less apparent motion, with an expected angular velocity of $\sim$\SI{1}{\degree/\year} at \SI{66}{\au}. The transformed images of the outer ring are not definitively more or less aligned after deshearing; the diffuse features are difficult to track.

\subsection{Phase Cross-Correlation\label{sec:phase_correlation}}

Although the desheared images in the previous section provide strong visual evidence for Keplerian disk motion, we need to quantify the motion to statistically verify if it is consistent with the local Keplerian velocity. Other studies have used models to fit clumps or profiles in the images \citep[e.g.][]{boccaletti_fast-moving_2015,ren_dynamical_2020}, but this requires an accurate model of the underlying intensity distributions of the features, and only uses a subset of the spatial information in the images.

Instead of fitting models to the surface brightness profiles, we utilized a cross-correlation technique. Phase cross-correlation is often used for image registration to estimate the linear displacement between images \citep{fienup_reconstruction_1978,fienup_phase_1982,guizar-sicairos_efficient_2008}. The basic principle is derived from the Fourier shift theorem-- let $\mathcal{F}$ be the Fourier transform operator
\begin{equation}
    \mathcal{F}\left\{f(x)\right\} \equiv F(\xi) =\int{f\left(x\right)\cdot e^{-i2\pi\xi x }dx }.
\end{equation}
The Fourier shift theorem states
\begin{equation}
    \mathcal{F}\left\{ f\left(x - \Delta x\right) \right\} = F(\xi)\cdot e^{-i2\pi\xi\Delta x}
\end{equation}
and the cross-power spectrum, $P$, of the Fourier transforms of both the shifted and original signals is
\begin{equation}
    \label{eqn:cross_prod_spec}
    P(\xi) = F(\xi)\left[F(\xi)\cdot e^{-i2\pi\Delta x \xi}\right]^* = \left|F\left(\xi\right)\right|^2\cdot e^{i2\pi\Delta x \xi}.
\end{equation}
The magnitude of $R$ is equivalent to the cross-power density of the two signals, and the phase has a linear ramp with a slope proportional to the shift. The inverse Fourier transform of \cref{eqn:cross_prod_spec} is the convolution of a Dirac-delta function with the autocorrelation of the input signal
\begin{align}
\begin{split}
    \label{eqn:phase_corr}
    \mathcal{F}^{-1}\left\{P(\xi)\right\} \equiv p(x) &= \int{\left|F\left(\xi\right)\right|^2\cdot e^{i2\pi\xi (x + \Delta x)}d\xi} \\
    &= \left[f \otimes f \right]\left( x \right) * \delta(x +\Delta x),
\end{split}
\end{align}
which can be simplified as the shifted autocorrelation function using the sifting property 
\begin{equation}
    \label{eqn:phase-correlation}
    p(x) = \left[f \otimes f\right]\left(x +  \Delta x \right).
\end{equation}
We refer to \cref{eqn:phase-correlation} as the phase correlogram when discretized and evaluated. 

Since the peak of the autocorrelation function is centered around zero, the displacement, $\Delta x$, can be estimated by fitting the location of the peak in the phase correlogram. The amplitude of the peak corresponds to the correlation of the two functions-- a higher amplitude indicates greater similarity between the signals, and a negative amplitude signifies a correlation between one signal and the negative of the other.

\begin{figure}
    \centering
    \script{plot_azimuthal_profiles_combo.py}
    \includegraphics[width=\columnwidth]{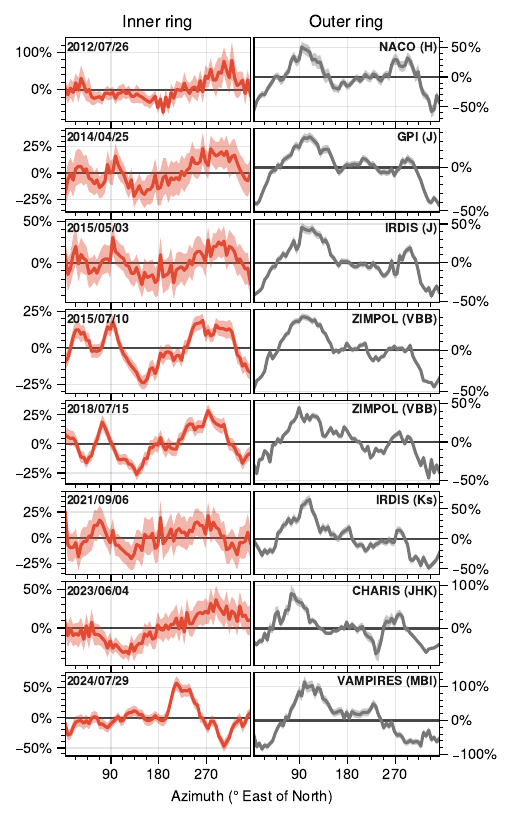}
    \caption{Azimuthal scattered-light profiles ($Q_\phi \times r^2$) for each dataset, normalized to the percent deviation from the mean value. The profiles were calculated using \ang{5}-wide azimuthal bins, the same as the polar images (\cref{fig:qphi_polar_inner}). The shaded areas are the 1$\sigma$ standard errors, using the $U_\phi \times r^2$ image as an error map. (Left) the inner ring, from \SIrange{15}{35}{\au}. (Right) the outer ring, from \SIrange{48}{110}{\au}.\label{fig:azimuthal_profiles}}
\end{figure}

We calculated the azimuthal profiles to use for phase correlation using the $Q_\phi \times r^2$ and $U_\phi \times r^2$ images. We convolved the images with a 1-pixel FWHM Gaussian kernel to reduce noise. The profiles were calculated from the $Q_\phi$ images, using the $U_\phi$ images as an error map. Each profile was normalized by subtracting the mean intensity and then dividing by the mean intensity to get the relative deviation from the mean. This preprocessing step eliminates flux units and centers the data around zero, preventing static offsets in the phase correlograms. The resulting azimuthal profiles for the inner and outer rings are presented in \cref{fig:azimuthal_profiles}.

We calculated the phase correlograms from azimuthal profiles for each combination of pairs of observations. There are 28 pair combinations from the 8 datasets, which are shown in \cref{fig:crosscorr_corner_inner,fig:crosscorr_corner_outer}. We estimated the measurement uncertainty in the phase correlograms with bootstrap resampling. We calculated the motion in degrees per year by dividing the displacement in the phase correlograms (in degrees) by the time between observations. Finally, we interpolated each pairwise phase correlogram onto a common angular velocity grid and calculated the mean phase correlogram, excluding combinations with aliasing issues (see \autoref{app:phase_correlograms}). The mean implicitly weighs combinations with high similarity more because the amplitude of phase correlograms is larger for more correlated signals.

\begin{figure}
    \centering
    \script{plot_azimuthal_profiles_crosscorr_combined.py}
    \includegraphics[width=\columnwidth]{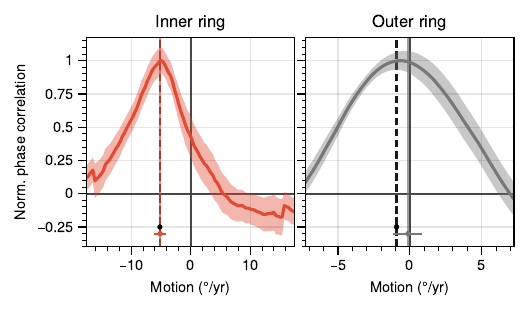}
    \caption{Mean phase correlograms for the inner and outer rings with $1\sigma$-uncertainty contours based on the azimuthal profiles in \cref{fig:azimuthal_profiles}. The locations of the peaks of the correlograms correspond to the estimated angular velocities of the azimuthal brightness profiles. The peak locations, measured with bootstrap resampling of the correlogram, are marked with vertical red and gray lines. The uncertainty in the peak estimates from bootstrap resampling are shown with a bar at the bottom of the plot. The expected Keplerian motion, based on the rings' separations, is marked with black vertical dashed lines. The motion measured via phase cross-correlation is consistent with Keplerian motion.\label{fig:azimuthal_profiles_crosscorr}}
\end{figure}

The mean phase correlogram is shown in \cref{fig:azimuthal_profiles_crosscorr} along with the expected Keplerian rotation rates for the disk regions. The angular velocity was measured from the location of the peak, using bootstrap resampling to estimate the uncertainties (\cref{tbl:corr_parameters}). This approach only considers the random error in the peak measurement, and not any systematic errors, so this uncertainty is a lower limit. The measured angular velocity of the inner ring is \SI{-5.14\pm0.9} {\degree/\year} and the expected Keplerian angular velocity is \SI{-5.18\pm0.18}{\degree/\year}, based on the uncertainty in the stellar mass and the location of peak intensity in the inner ring. The negative value indicates clockwise motion projected on the sky, consistent with previous observations \citep{ligi_investigation_2018,gratton_blobs_2019,garg_kinematic_2022,hammond_confirmation_2023}. The outer ring has an estimated angular velocity of \SI{-0.13\pm0.93}{\degree/\year} and an expected Keplerian velocity of \SI{-0.93\pm0.08}{\degree\year}.

We assessed the statistical significance of the measured angular velocity of brightness features within each ring  using an unpaired Student's t-test with three hypotheses: (1) no motion (i.e., the features are stationary on the sky), (2) motion consistent with local Keplerian rotation at the ring's separation, and (3) motion consistent with the Keplerian velocity of companion b at \SI{37.2}{\au}. For the inner ring, the measured angular velocity deviates from the no-motion hypothesis at the $5.7\sigma$ level, is consistent with local Keplerian motion at the $0.05\sigma$ level, and differs from the angular velocity of companion b by $3.3\sigma$. These results support the interpretation that the inner ring features are not influenced by any companions interior or exterior to the ring. In contrast, the outer ring's angular velocity deviates $0.1\sigma$ from the no-motion hypothesis, $0.8\sigma$ from local Keplerian motion, and by $2.2\sigma$ from the velocity of companion b. We cannot conclusively determine the outer ring's dynamics from these results, but they show that the outer ring features are not likely to be driven by companion b.

\begin{deluxetable*}{lccc}
    \centering
    \tablecaption{HD 169142 Keplerian angular velocities.\label{tbl:corr_parameters}}
    \tablehead{
        \colhead{Name} &
        \colhead{Inner ring} &
        \colhead{HD 169142 b$^\dagger$} & 
        \colhead{Outer ring}
    }
    \startdata
    $r_\text{max}$ (\si{\au}) & \num{20.9\pm0.4} & \num{37.2\pm1.5} & \num{65.9\pm3.8}\\
    \hline
    Keplerian $\omega$ (\si{\degree/\year}) & \num{-5.18\pm0.18} & \num{-2.18\pm0.14} & \num{-0.93\pm0.08} \\
    Measured $\omega$ (\si{\degree/\year}) & \num{-5.14\pm0.9} & & \num{-0.13\pm0.93} \\
    \enddata
    \tablecomments{$\omega$: angular velocity. The expected values assume a stellar mass of \SI{1.89\pm0.08}{\Msun}.}
    \tablerefs{($\dagger$): \citealt{hammond_confirmation_2023}}
\end{deluxetable*}

%% file: 06_compare_to_alma.tex
\section{Comparing Scattered-light to \SI{1.3}{\milli\meter} Dust Emission\label{sec:compare_to_alma}}

\begin{figure}
    \centering
    \includegraphics[width=\columnwidth]{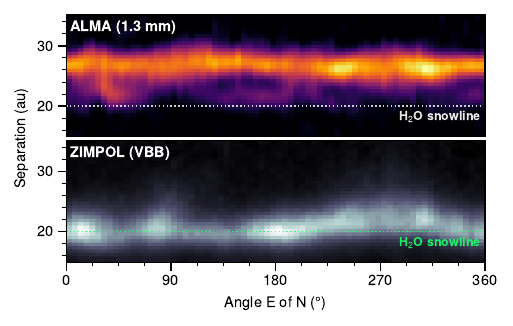}
    \caption{Polar images of the \formatdate{15}{7}{2018} ZIMPOL scattered-light ($Q_\phi \times r^2$) and \SI{1.3}{\milli\meter} dust emission for the inner ring (\SIrange{15}{35}{\au}). The ZIMPOL data has been desheared to correct for Keplerian motion between the two observation dates. The \ch{H2O} snow line transition region is marked with a horizontal dashed line in both panels \citep{booth_tracing_2023}.\label{fig:mean_polar_alma_inner}}
\end{figure}

\begin{figure}
    \centering
    \includegraphics[width=\columnwidth]{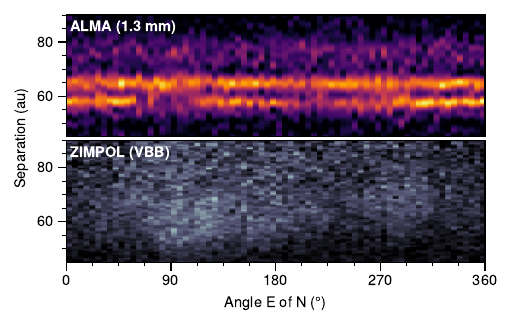}
    \caption{The same as \cref{fig:mean_polar_alma_inner}, but for the outer ring from \SI{45}{\au} to \SI{90}{\au}.\label{fig:mean_polar_alma_outer}}
\end{figure}

Under the assumption that both the inner and outer ring features move with Keplerian motion, we can directly compare the non-axisymmetric features in the \SI{1.3}{\milli\meter} dust continuum emission with the scattered-light images. We used our Keplerian deshearing process to transform the 2018 ZIMPOL data to the date of the ALMA \SI{1.3}{\milli\meter} observations. The ZIMPOL data has the highest angular resolution of all our scattered light images, and the 2018 epoch is closest in time to the ALMA observations. We show a comparison between the ALMA and ZIMPOL data for the inner and outer rings in \cref{fig:mean_polar_alma_inner,fig:mean_polar_alma_outer}, respectively.

The polar image of the millimeter dust emission shows a prominent inner ring around \SI{26}{\au} with non-axisymmetric features. Interior to that ring are various tendrils of emission--these features do not appear consistent with spiral arms, but appear more like a broken ring. We observed that the scattered light traces the edges of the millimeter dust emission: the depression in intensity and separation in scattered light at position angle \ang{45} (S1) fits like a puzzle piece to one of the tendrils in the ALMA continuum image. A similar effect can be seen from \ang{100} to \ang{220} (S2). The outward bowing around \ang{270} in the scattered light pieces together well with the ALMA continuum image, where there is a lack of tendrils interior to the ring. Globally, the scattered light appears interior to the millimeter dust emission, as shown in the radial profiles (\cref{fig:radial_profiles}), which may indicate a leaky dust trap \citep{taki_dust_2016,stammler_leaky_2023,huang_leaky_2025}.

The outer ring has fewer non-axisymmetric features in the dust continuum: the three dust rings have eccentric shapes ($e\sim0.09$; \citealp{perez_dust_2019}) and intensity dips around \ang{90} and \ang{200}. The \ang{90} dip in continuum emission lines up with the bright feature in scattered light (\cref{fig:qphi_polar_outer}). \citet{hammond_confirmation_2023} suggests that this bright feature corresponds to a spiral wake trailing the candidate protoplanet HD 169142 b at \SI{37}{\au}, which aligns with the primary spiral associated with blob D in \citet{gratton_blobs_2019}. However, in this case, the ring would be expected to brighten due to an increase in density rather than exhibit a localized decrease in intensity.

%% file: 07_discussions.tex
\section{Discussion\label{sec:discussion}}

The observed scattered-light and dust continuum structures in HD 169142 reveal a complex disk morphology, including concentric rings, azimuthal asymmetries, and potential signatures of planet-disk interactions. By analyzing the spatial distribution of polarized light and comparing it with dust emission, we identify differences in the radial and azimuthal profiles that provide insight into the underlying disk dynamics. In this section, we discuss the implications of these features in the context of disk evolution, planet formation, and previously proposed models for HD 169142.

\subsection{Effects of Scattering Geometry\label{sec:scattering}}

Scattered-light images are sensitive to changes in the angle between the star, disk, and observer-- for example, in an inclined disk the forward-scattering light should appear brighter than the backward-scattering light \citep[e.g.,][]{milli_optical_2019}. Furthermore, polarimetric data are also affected by the scattering angle and dust properties, which affect the fractional polarization of the scattered light. Such brightness asymmetries have global effects on disk illumination and are oriented with the disk major and minor axes. Because these asymmetries are due to the projected geometry of the disk on the sky, they do not move or change over time.

The brightness features of the inner ring of HD 169142 do not match this description: the features are highly localized, not aligned with the major or minor axes, and move over time. Furthermore, since the disk has a low inclination, the scattering angles are similar across the image. The total peak-to-valley variance in the partial polarization (assuming Rayleigh scattering) is only 9\% for a disk with \ang{12.5} of inclination--however, we see deviations on the scale of 10\% to 30\% in the azimuthal profiles (\cref{fig:azimuthal_profiles}). This evidence suggests that the brightness features in the inner ring are not a result of scattering geometry, but are physical features of the disk.

\subsection{Shadows\label{sec:shadows}}

A companion, circumplanetary disk, or a clumpy or warped disk inside the inner cavity of HD 169142 could be casting shadows on the inner ring, creating dips in the azimuthal brightness. \citet{bertrang_hd_2018,bertrang_moving_2020} reported a narrow shadow moving across the \SIrange{0}{90}{\degree} quadrant over a five-year baseline (S1 from \cref{fig:qphi_polar_inner_annotated}). They reported super-Keplerian motion implying evidence for a \SI{\sim13}{\au} companion with a mass of approximately \SI{7}{\Mjup}, based on the size of the Hill sphere needed to cast the shadow. However, our analysis does not support this interpretation, as demonstrated by both the visual alignment of the same shadow feature in Keplerian-corrected polar images (\cref{fig:qphi_polar_inner}) and the phase correlation analysis presented in \cref{sec:phase_correlation}.

A misaligned inner disk with respect to the outer disk (also known as a broken or warped disk) can create shadows \citep[e.g.][]{marino_shadows_2015,stolker_shadows_2016,avenhaus_exploring_2017,benisty_shadows_2018,pinilla_variable_2018}, and the motion of these shadows is associated with the inner disk precession--which will be different than the inner ring's Keplerian velocity. While there are measurements of the morphology and orientation of HD 169142's inner disk \citep{bohn_probing_2022}, estimating precession rates requires detailed knowledge of torques affecting the disk, which is beyond the scope of this work. Shadows in an almost face-on outer disk, cast by a warped inner disk, form in pairs $\sim$\ang{180} apart and can appear narrow or broad, depending on the misinclination \citep[e.g.,][]{stolker_shadows_2016,benisty_shadows_2017,ginski_disk_2021}. We do not observe any shadows in our data that exhibit the expected \ang{180} morphology, and we do not pursue this hypothesis further.

We consider that the shadow features we see are aliased, caused by an object with a significantly faster angular frequency. However, in this study, we have eight epochs that span twelve years with irregular spacing, which makes it difficult for aliasing to work. For example, the shortest baseline between data is the 2015 IRDIS epoch and the 2015 ZIMPOL epoch which are 64 days apart. The motion we measured over this period was \SI{-0.9 \pm 0.16}{\degree}. In an aliasing scenario, the features could have moved \ang{-360.9}, which corresponds to an angular velocity of \SI{-2059.7 \pm 0.9}{\degree/\year}. If that is the case, the shadow would have shifted by \ang{~285} between the \formatdate{25}{4}{2014} and \formatdate{3}{5}{2015} epochs in addition to completing five full revolutions. This is clearly not seen in the data. The fact that features remain well aligned across multiple epochs therefore strongly supports the simple interpretation that the measured velocities represent the true, slower motion of the shadows.

\subsection{Annular Ring Gaps}

Massive companions can sculpt out annular gaps in disks through accretion of material \citep{goodman_planetary_2001,wolf_observability_2005}. Empirical power-law relationships have been found between the width and depth (contrast) of a gap in scattered light and the companion mass and disk viscosity \citep{dong_what_2017}. We calculated the estimated mass of a gap-clearing companion following the same procedure as \citet{dong_what_2017}. The gap parameters (location, width, and depth) were calculated from the surface-brightness profile, and we applied their empirical relationships to estimate the planet-star mass ratio for three disk viscosities ($\alpha=10^{\mathrm{-}4}$,$10^{\mathrm{-}3}$, and $10^{\mathrm{-}2}$). Our results are listed in \cref{tbl:ring_parameters}. The resulting planet masses range from \SIrange{0.2}{2.2}{\Mjup}, consistent with the findings of \citet{dong_what_2017}. Using the Keck/NIRC2 L` mass sensitivities as an upper limit for companion mass (\SIrange{1.4}{2.2}{\Mjup}), the implied disk viscosity is less than \numrange{4e-3}{1e-2}, respectively, which is reasonable \citep{rafikov_protoplanetary_2017}.

The ring gap has also been studied using optically thick tracers with ALMA. Velocity maps of \ch{CO} emission have been employed to characterize the global disk geometry and extent \citep{fedele_alma_2017,yu_mapping_2021,bohn_probing_2022,garg_kinematic_2022}. These velocity profiles predominantly exhibit Keplerian rotation, with a global pressure gradient that is well described by a tapered disk model. Subtracting this global model reveals localized excess pressure within the D2 gap, approximately coincident with the \SI{37}{\au} candidate companion \citep{yu_mapping_2021,garg_kinematic_2022}. Additionally, excess emission from \ch{SO} and \ch{SiS} volatiles has been detected at the same location as the \ch{CO} kinematic excess, suggesting a circumplanetary-disk-driven gas outflow \citep{law_so_2023,keyte_volatile_2024}. While these findings provide intriguing evidence for planetary influence, they remain indirect and rely on chemical modeling of disk environments. Furthermore, the observations have limited angular resolution (FWHM$\sim$\ang{;;0.2}). Alternative explanations for the molecular abundance enhancements are local variations in the gas-to-dust ratio or radiation--perhaps caused by shadowing from the inner ring.

Simulations by \citet{perez_dust_2019} suggest that the smaller ring gaps within the outer dust rings can be explained by the combined influence of a mini-Neptune and a Jovian-mass planet. The asymmetric ring sizes between B2 to B4 and the varying gap widths of D3 and D4 are more consistent with a migrating mini-Neptune at \SI{64}{\au}--coincident with the B3 ring--than with a companion in each gap or gaps carved by a single, non-migrating planet. Additionally, \citet{perez_dust_2019} demonstrated that adding a giant planet within the D2 gap can account for the observed eccentricity of the B2 and B3 rings ($e\sim0.09$). The low mass of the proposed outer companion would not produce detectable features in our scattered-light images, such as localized emission, gaps, or spiral structures, nor would it be directly detectable. Consequently, we are unable to place meaningful observational constraints on its presence or absence. 

\subsection{Spirals\label{sec:spirals}}

A Jovian-mass companion is expected to generate prominent spiral arms observable in both scattered light and dust emission \citep{goodman_planetary_2001,wolf_observability_2005,dong_how_2017}. While no direct evidence of spiral structures is found in dust emission maps \citep{perez_dust_2019}, some studies have reported spiral-like features in scattered-light observations \citep{gratton_blobs_2019,hammond_confirmation_2023}. In particular, \citet{hammond_confirmation_2023} labels the bright feature in the outer ring around PA$\sim$\SIrange{90}{130}{\degree} in the 2015 IRDIS data as a spiral wake. While we see this bright clump in many of our datasets (\cref{fig:qphi_polar_outer}), the 2017 ALMA data actually shows a dimming of \SI{1.3}{\milli\meter} emission at the same location (\cref{fig:mean_polar_alma_outer}), which weakens the spiral-wake theory; planet-induced spirals should be at least as bright or brighter at millimeter wavelengths compared to the disk background \citep{dong__m_2015,rosotti_spiral_2020,speedie_observing_2022}.

Furthermore, spiral arms induced by a companion rotate at the same speed as the companion, even if they extend to separations where the local Keplerian velocity is different \citep{dong_how_2017,dong_eccentric_2018,ren_dynamical_2020}. Our dynamical analysis of azimuthal features within both the inner and outer rings indicates motion consistent with local Keplerian rotation (\autoref{sec:dynamical_analysis}). Therefore, it is unlikely for the bright features in the outer ring to be a spiral wake as reported by \citet{hammond_confirmation_2023}, or for the inner ring's features to be associated with spirals as reported by \citet{gratton_blobs_2019}.

\subsection{Vortices\label{sec:vortices}}

Vortices in disks can be local dust traps and appear as large, bright arcs \citep{fung_cooling-induced_2021,ma_vortex-induced_2025}. These vortices can be caused by perturbations in the disk such as planet-driven spirals \citep{cimerman_planet-driven_2021,cimerman_emergence_2023}, embedded planets \citep{val-borro_vortex_2007}, or snow lines. The overdensities observed in the inner and outer rings could potentially be emerging vortices driven by Rossby wave instability \citep{val-borro_vortex_2007,meheut_dust-trapping_2012,richard_vortex_2016}. These vortices evolve rapidly, initially manifesting as many small clumps before merging into a single large arc after only a few hundred orbits \citep{ma_vortex-induced_2025}. To explain the multiple features in the inner ring, this would suggest a planet in its earliest stages of formation--around 100 orbits at \SI{37}{\au}, which equates to $\sim10^4$ \si{\year}.

While this is an intriguing interpretation, it is statistically unlikely to observe a phenomenon with a time scale $\sim10^4$ \si{\year} compared to the system age of \SI{7.4}{\mega\year}. Furthermore, such a protoplanet would be bright, and the lack of prominent signal in the thermal-infrared observations \citep{reggiani_discovery_2014,biller_enigmatic_2014,wallack_survey_2024} contradicts this scenario. Moreover, the bright feature observed in the outer ring in scattered light coincides with a dimming in the millimeter dust emission (\cref{fig:mean_polar_alma_outer}), which suggests that a vortex alone is not a good explanation because there should be a higher density of millimeter grains.

%% file: 08_conclusions.tex
\section{Conclusions\label{sec:conclusions}}

In this work, we presented polarimetric images of the transition disk HD 169142 over the span of twelve years, analyzing the dynamics of long-lived, non-axisymmetric features.
\begin{itemize}
    \item[-] We highlighted features in the inner and outer rings that are consistent across visible to near-IR wavelengths and are long-lived over the 12-year baseline (\cref{fig:qphi_mosaic,fig:qphi_polar_inner,fig:qphi_polar_outer}). 
    \item[-] We measured the azimuthal brightness of both the inner and outer rings in scattered light. The inner ring varies 10\% to 30\% compared to the mean flux and the outer ring varies 50\% compared to its mean (\cref{fig:azimuthal_profiles}).
    \item[-] We analyzed the motion of the rings using phase cross-correlation and concluded the brightness features of the inner ring are consistent with the local Keplerian velocity (\cref{fig:azimuthal_profiles_crosscorr,tbl:corr_parameters}). This is contrary to interpretations claiming they are spirals induced by a companion in the large gap (\autoref{sec:spirals}) or shadows from an inner companion or disk (\autoref{sec:shadows}).
    \item[-] We did not conclusively determine the origin of the motion of the outer ring, obtaining results moderately consistent with Keplerian motion (\cref{fig:azimuthal_profiles_crosscorr,tbl:corr_parameters}). Longer time baselines are necessary to improve the precision of these claims with our methods.
    \item[-] We compared the scattered-light features to the \SI{1.3}{\milli\meter} dust continuum emission and found strong correlations. The scattered light appears to trace the inner edge of the ring of large dust grains, indicative of complex disk interactions and a leaky dust trap (\autoref{sec:compare_to_alma}, \cref{fig:mean_polar_alma_inner}).
    \item[-] We see no evidence for planet-disk interactions in the form of spiral arms, shadows, or vortices caused by the candidate Jovian-mass protoplanet within the annular gap (\autoref{sec:discussion}). These features may still exist but are either too subtle to detect or obscured by other disk phenomena.
\end{itemize}

This study represents one of only a few multi-epoch dynamical analyses of circumstellar disks to date \citep[e.g.,][]{boccaletti_fast-moving_2015,debes_chasing_2017,stolker_variable_2017,ren_decade_2018,sallum_systematic_2023,skaf__2023,avsar_search_2024}. This analysis is enabled by the exceptional angular resolution, sensitivity, and temporal baseline afforded by mature high-contrast polarimetric instrumentation and the growing wealth of archival data. By combining multi-wavelength observations across epochs, we revealed the evolving morphology of disk substructures and placed constraints on the origin of their underlying dynamics. While polarimetric surveys have identified a large number of protoplanetary disks, these findings highlight the unique scientific value of targeted, long-term follow-up, which allows for detailed characterization of disk dynamics and planet-disk interactions.

Future high-resolution ALMA continuum observations will also be important to track the evolution of this system's features. Determining whether the azimuthal features are long-lived and how they move within the disk will further constrain the interpretations of the features, especially if they are related to any embedded protoplanets via complex disk interactions. Contemporaneous scattered-light images would aid in detecting motion in the outer ring, increasing the length of the baseline.

Space-based observations with JWST make use of the cold and stable space environment for high-contrast near- and mid-infrared imaging. These modes are well-suited to studying HD 169142, in theory, but are hard to realize in practice. JWST would be sensitive to a forming protoplanet, but the expected separation (\SI{37}{\au}; \ang{;;0.3}) is covered by the MIRI focal plane masks \citep{boccaletti_jwstmiri_2022} and challenging for NIRCAM, requiring use of the small bar mask \citep{girard_jwstnircam_2022}.

Future observations with extremely large telescopes will be pivotal for answering questions about planet-forming disks. The improved sensitivity and angular resolution will provide stronger constraints on photospheric detections of protoplanets and circumplanetary disks, as well as the detection of fainter disk features associated with lower planetary masses. Additionally, these capabilities will probe a new demographic regime, including younger systems (which are more obscured with dust and fainter) and systems in more distant star-forming regions (with lower apparent magnitudes and more severe angular resolution constraints).

%% file: acknowledgements.tex
\begin{acknowledgements}
We wish to recognize and acknowledge the significant cultural role and reverence that the summit of Maunakea has always had within the Indigenous Hawaiian community. We are grateful and thank the community for the privilege of conducting observations from this sacred mauna.

We thank telescope operators L. Schumacher, A. Walk, and E. Dailey for their support during the observations used in this work. We thank T. Tobin for extracting the CHARIS data with the \texttt{ADEPTS} pipeline. We thank G. H. Bertrang for sharing the ZIMPOL data used in this work.

This research was funded by the Heising-Simons Foundation through grant \#2020-1823. M. Bensity has received funding from the European Research Council (ERC) under the European Union's Horizon 2020 research and innovation program (PROTOPLANETS, grant agreement No. 101002188). S.P. acknowledges support from Fondecyt Regular 1231663 and ANID ---Millennium Science Initiative Program--- Center Code NCN2024\_001. M. F. has received funding from the European Research Council (ERC) under the European Unions Horizon 2020 research and innovation program (grant agreement No. 757957).

Based on data collected at Subaru Telescope, which is operated by the National Astronomical Observatory of Japan. The development of SCExAO is supported by the Japan Society for the Promotion of Science (Grant-in-Aid for Research \#23340051, \#26220704, \#23103002, \#19H00703, \#19H00695, and \#21H04998), the Subaru Telescope, the National Astronomical Observatory of Japan, the Astrobiology Center of the National Institutes of Natural Sciences, Japan, the Mt Cuba Foundation and the Heising-Simons Foundation. 
\end{acknowledgements}

%% file: appendix_01_crosscors.tex
\section{Phase Correlograms\label{app:phase_correlograms}}

\begin{figure*}
    \centering
    \includegraphics[width=\textwidth]{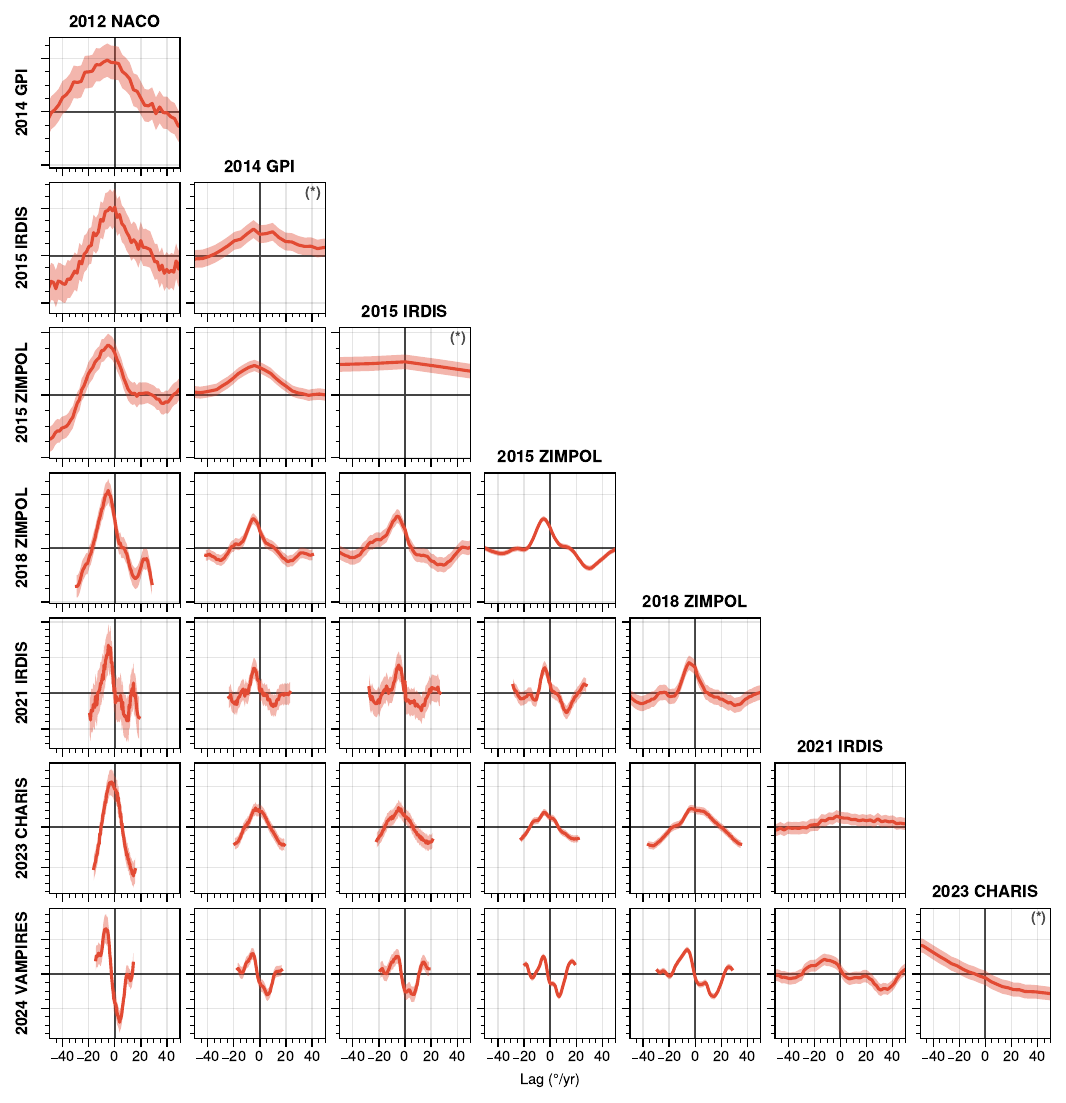}
    \caption{Phase corellograms for pairs of azimuthal profiles of the inner ring (\SIrange{15}{35}{\au}). The row and column labels signify the two epochs correlated in the subsequent correlogram. Plots with a (*) are excluded from the mean combination and analysis in \cref{fig:azimuthal_profiles_crosscorr}.\label{fig:crosscorr_corner_inner}}
\end{figure*}

\begin{figure*}
    \centering
    \includegraphics[width=\textwidth]{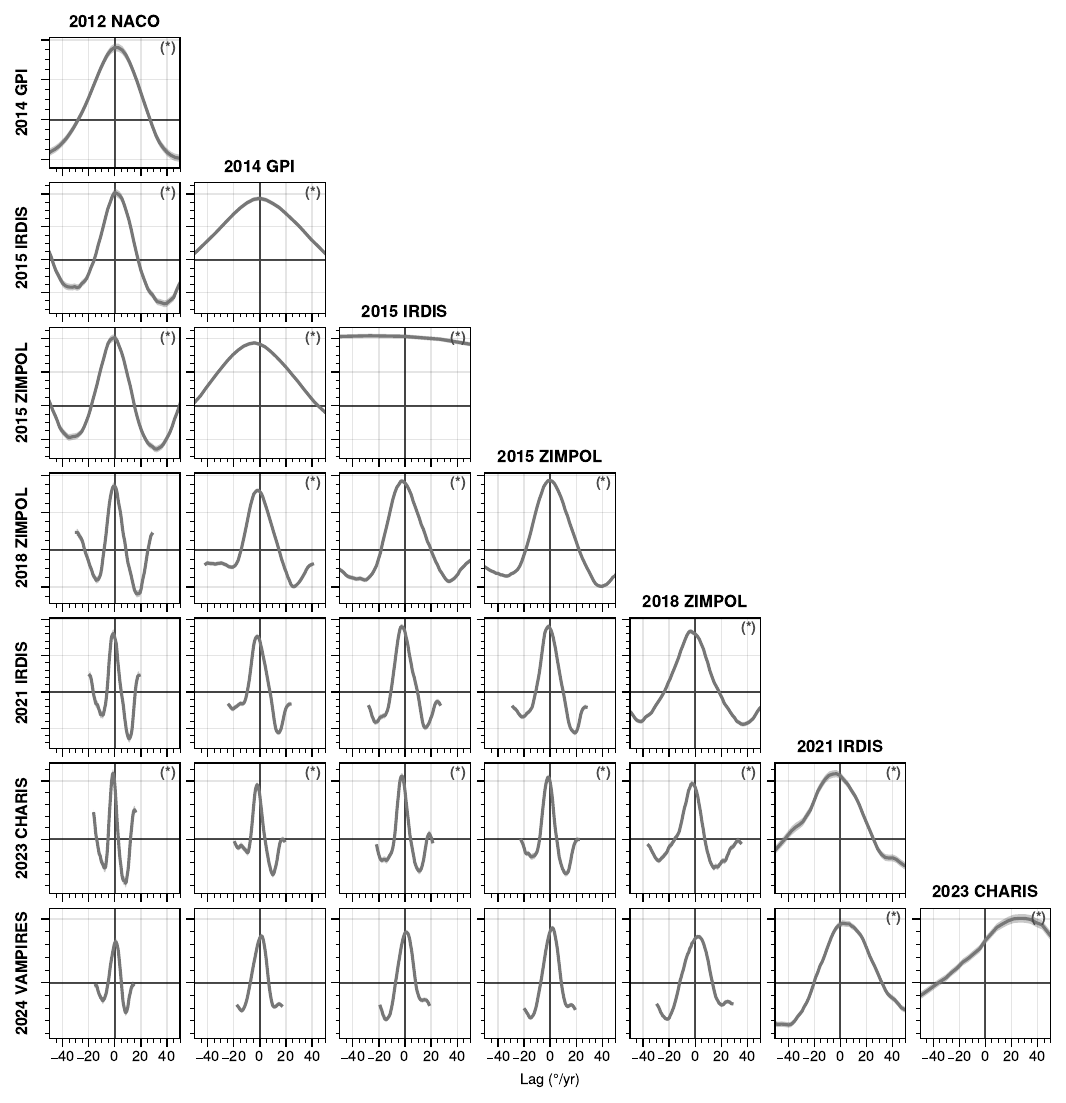}
    \caption{The same as \cref{fig:crosscorr_corner_inner}, but for the outer ring (\SIrange{48}{115}{\au}).\label{fig:crosscorr_corner_outer}}
\end{figure*}

We performed phase cross-correlation on pairs of azimuthal profiles as outlined in \autoref{sec:phase_correlation}. The resulting pairwise correlograms are shown in \cref{fig:crosscorr_corner_inner,fig:crosscorr_corner_outer}, which were used to generate the mean correlogram in \cref{fig:azimuthal_profiles_crosscorr}. Some of the correlograms, such as those from the 2015 ZIMPOL and 2015 IRDIS observations, exhibit strong aliasing effects. This is due to the finite sample size of the azimuthal profiles (\ang{5} bins) and the short time intervals between epochs.

When the apparent motion between epochs is smaller than the bin size, aliasing artifacts arise. To mitigate this, pairs of epochs with less apparent motion than the bin size, assuming Keplerian rotation of the rings, were excluded from the mean combination in \cref{fig:azimuthal_profiles_crosscorr}. These pairs are marked with a ``(*)'' in \cref{fig:crosscorr_corner_inner,fig:crosscorr_corner_outer}. Although this aliasing could be addressed using sub-pixel interpolation methods, we leave this for future work, as the long baselines in our dataset still allow for meaningful dynamic analysis, even for the outer ring.

%% file: appendix_02_data.tex
\section{Data and Code Availability\label{app:data_availability}}

This study used the reproducibility software \texttt{showyourwork} \citep{luger_mapping_2021}, which leverages continuous integration to create the figures and compile this manuscript programmatically. The git repository associated with this study is publicly available at \url{https://github.com/mileslucas/HD169142_disk_paper}. The reduced SCExAO data are available at \url{https://doi.org/10.5281/zenodo.17089001}. Raw observational data for the SCExAO observations are available from the SMOKA archive\footnote{\url{https://smoka.nao.ac.jp/}}. The GPI data were obtained from \citet{rich_vizier_2022} and the NACO data were obtained from \citet{de_regt_vizier_2024}. All remaining data are available upon reasonable request.